%% file: mana050802v2-induction.tex
\RequirePackage{txfonts}\normalfont     

\documentclass[11pt,
onecolumn,oneside,a4paper,article
,italian,frenchb,german,
swedish,british%
]{memoir}

\newcommand{\usetxfonts}{\usepackage{txfonts}}
\input{preamblemem.tex}

\alleugreek

\usepackage{paralist}

\usepackage[hypertex,breaklinks,pdfauthor={P G L  Porta Mana}]{hyperref}

\usepackage[numbers,sort&compress]{natbib}
\usepackage{memhfixc}

\setlxvchars
\setxlvchars

\reparticle
\checkandfixthelayout
\fixpdflayout

\copypagestyle{manaart}{plain}
\makeheadrule{manaart}{\headwidth}{\normalrulethickness}
\makeoddhead{manaart}{\small\textsc{Porta Mana, M\aa{}nsson,
  Bj\"ork}}{}{\small\textit{The Laplace-Jaynes approach to induction}}

\newenvironment{acknowledgements}{\chapter*{Acknowledgements}\addcontentsline{toc}{chapter}{Acknowledgements}}{\par}

\theoremstyle{plain}
\newtheorem*{corol}{Proposition}

\newcommand{\tprod}{\mathop{\textstyle\prod}\limits}
\newcommand{\tsum}{\mathop{\textstyle\sum}\limits}
\newcommand{\tland}{\mathop{\textstyle\bigwedge}\limits}
\newcommand{\tlor}{\mathop{\textstyle\bigvee}\limits}

\newcommand{\yn}{n}
\newcommand{\yN}{N}
\newcommand{\yNf}{\Bar{\yN}}
\newcommand{\yL}{L}
\newcommand{\yLf}{\Bar{\yL}}
\newcommand{\yM}{M}
\newcommand{\yff}{f}
\newcommand{\yf}{\Bar{\yff}}

\newcommand{\yj}{\sigma}
\newcommand{\yo}{\tau}
\newcommand{\yg}{Q}

\newcommand{\yme}{\mathfrak{M}}

\newcommand{\yI}{I}
\newcommand{\yIs}{\yI_\text{s}}
\newcommand{\yIu}{\yI_\text{u}}
\newcommand{\yE}{E}

\newcommand{\yR}{R}
\newcommand{\yD}{D}

\newcommand{\yk}{K}

\newcommand{\yQ}{\gamma}
\newcommand{\yP}{\varGamma}
\newcommand{\yqq}{q}
\newcommand{\yq}{\Bar{\yqq}}
\newcommand{\yqc}{{\yq}\sptilde}
\newcommand{\yupp}{\yqq}
\newcommand{\yuep}{\Hat{\yupp}}
\newcommand{\yup}{\Bar{\yupp}}

\newcommand{\ysim}{\varDelta}

\newcommand{\yC}{C}
\newcommand{\yS}{S}

\newcommand{\yl}{\lambda}

\newcommand{\yv}{v}
\newcommand{\yal}{\alpha}
\newcommand{\yvo}{v_0}
\newcommand{\yvu}{v_1}
\newcommand{\yCmb}{\yC_\text{MB}}
\newcommand{\yCpu}{\yC_\text{P}}
\newcommand{\yCne}{\yC_\text{NE}}

\pretitle{\begin{center}\LARGE\bfseries}
\predate{\begin{center}(Dated: }
\postdate{)\end{center}}

\providecommand{\affiliation}[1]{\textit{\small#1}}
\providecommand{\pacs}[1]{{\small\textsc{PACS} numbers: #1}}
\providecommand{\msc}[1]{{\small\textsc{MSC} numbers: #1}}
\providecommand{\email}[1]{\texttt{\href{mailto:#1}{#1}}}

\title{The Laplace-Jaynes approach to induction
\\[3\jot]
\textnormal{\large Being part II of\\``From `plausibilities of
    plausibilities' to state-assignment methods''}}

\author{\firstname{P. G. L.} \surname{Porta Mana},\thanks{Email: \email{mana@kth.se}}
\quad
\firstname{A.} \surname{M{\aa}nsson},
\quad
\firstname{G.} \surname{Bj\"{o}rk}
\\
\affiliation{Kungliga Tekniska H\"ogskolan, Isafjordsgatan
  22, SE-164\,40 Stockholm, Sweden}
}

\date{29 April 2007}

\begin{document}
\bibliographystyle{apsrevmananum} 

\setlength{\droptitle}{-3\onelineskip}

\maketitle
\begin{abstract}
  An approach to induction is presented, based on the idea
  of analysing the context of a given problem into
  `circumstances'. This approach, fully Bayesian in form
  and meaning, provides a complement or in some cases an
  alternative to that based on de~Finetti's representation
  theorem and on the notion of infinite exchangeability.
  In particular, it gives an alternative interpretation of
  those formulae that apparently involve `unknown
  probabilities' or `propensities'. Various advantages and
  applications of the presented approach are discussed,
  especially in comparison to that based on
  exchangeability. Generalisations are also discussed.
  \\[2\jot]
  \pacs{02.50.Cw,02.50.Tt,01.70.+w}\\
  \msc{03B48,60G09,60A05}
\end{abstract}


\setlength{\epigraphwidth}{.6\columnwidth}
\epigraph{Note, to head off a common misconception, that
  this is in no way to introduce a ``probability of a
  probability''. It is simply convenient to index our
  hypotheses by parameters [\ldots] chosen to be
  numerically equal to the probabilities assigned by those
  hypotheses; this avoids a doubling of our notation. We
  could easily restate everything so that the
  misconception could not arise; it would only be rather
  clumsy notationally and tedious verbally.}%
{\textsc{E. T. Jaynes}, \textit{Monkeys, kangaroos, and
    $N$}~\citep{jaynes1986d_r1996}, p.~12}

\chapter{Dramatis  personae, notatio, atque philosophia}
\label{sec:personae}

We continue here our exploration of the notion of
`circumstance' and of its applications in plausibility
theory. Through this notion we shall here approach no less
than the question of induction, \ie, of the prediction of
unobserved evens from knowledge of observed (similar)
ones.
The study can be read independently of the previous
one~\citep{portamanaetal2006}, although the two elucidate
each other.

The following characters appear in this study, the first
two having a walk-on part only:

\emph{The `propensitor':} a scholar who conceives of a `physical
disposition' of some phenomena to occur with definite
relative frequencies, and call this physical
characteristic `propensity'. This scholar sometimes
makes inferences about propensities by means of
probability as degree of belief.

\emph{The `frequentist':} a scholar who conceives probability as
`limit relative frequency' of a series of `trials' (or
something like that). Frequentists are often propensitors
at heart.

\emph{The `de~Finettian':} a scholar who conceives
probability as `degree of belief', pedantically insisting
on the subjective nature of all plausibility assignments,
and through the notion of infinite exchangeability makes
sense of the frequentist's and propensitor's
voo\-doo\-is\-tic practises. The views and tools of this
scholar~\citep[\eg,][]{definetti1931,definetti1937_t1964,hewittetal1955,definetti1970_t1990,definetti1970b_t1990,heathetal1976,diaconis1977,diaconisetal1980,bernardoetal1994}
are supposed known to the reader.

\emph{The `plausibilist' or `probability logician',} who in
this study will be your humble narrator: a scholar who
conceives of probability as a formalisation and
schematisation of the everyday notions of `plausibility'
and `probability' --- just as truth in formal logic is
such a formalisation and schematisation of the everyday
notion of `truth' --- and does not dwell more than so in
its meaning. The views of this scholar are a distillate of
the philosophies and views and/or the works of
Laplace~\citep{laplace1812_r1820},
Johnson~\citep{johnson1924},
Jeffreys~\citep{jeffreys1939_r1998,jeffreys1931_r1957,jeffreys1955},
Cox~\citep{cox1946,cox1961},
Jaynes~\citep{jaynes1954,jaynes1959,jaynes1994_r2003},
Tribus~\citep{tribus1969},
de~Finetti~\citep{definetti1970_t1990,definetti1970b_t1990},
Adams~\citep{adams1975,adams1998},
Hailperin~\citep{hailperin1996}, and
others~\citep[\eg,][]{keynes1921,ramsey1926_r1950,ramsey1931_r1950,kolmogorov1933_t1956,koopman1940,koopman1940b,koopman1941,polya1954,polya1954b_r1968,los1955,gaifman1964,scottetal1966,gaifmanetal1982,bernardoetal1994,gaifman2003,gregory2005}.


Plausibilists agree with de~Finettians on basically all
points. The only difference is in emphasis: although they
recognise the `subjective' character of initial
plausibility assignments --- \ie, that these are matter of
convention ---, they also think that it is not such a big
deal. It should in any case be no concern of plausibility
theory, which is `impersonal in the same sense as in
ordinary logic: that different people starting from the
same [assignments] would get the same answers if they
followed the rules'~(Jeffreys~\cite{jeffreys1955},
reinterpreted). This will surely be self-evident when
plausibility theory will reach its maturity, just as it is
self-evident in formal logic today. In fact, the same
subjective, conventional character is present --- no more,
no less --- in formal logic as regards the initial truth
assignments, those which are usually called `axioms' or
`postulates'. Albeit in formal logic there is perhaps less
place for disagreement amongst people about initial truth
assignments, the possible choice being there between
`true' and `false', or `0' and `1' only --- and not
amongst a continuum of possibilities $\clcl{0,1}$ as in
plausibility theory.

The above remark is not meant to diminish the historical
importance of de~Finetti's Nietzschean work in
plausibility theory. Emphasis on subjectivity may still be
necessary sometimes. In this study we shall also use terms
like `judgement' to this purpose.


The following notation will be used:

$\pr(A \cond B)$ denotes the plausibility of the
proposition $A$ conditional on $B$ or, as we shall also
say, given $B$ or in the \emph{context} $B$. The
proposition $B$ is supposed to express all knowledge
(including beliefs), data, and `working hypotheses' on the
grounds of which we make the plausibility assignment for
$A$. We are intransigent as regards the necessity and
appropriateness of always specifying the context of a
plausibility (as well as that of a truth!),\footnote{The
  context may also have a clarifying r\^ole in formal
  logic. \Cf\ \eg\ the studies by
  Adams~\cite{adams1975,adams1988,adams1998},
  Lewis~\cite{lewis1976,lewis1986},
  Hailperin~\cite{hailperin1984,hailperin1996},
  Barwise~\cite{barwise1989,barwise1985}, and
  Gaifman~\cite{gaifman2002}.} a point also stressed by
Jeffreys~\citep{jeffreys1931_r1957,jeffreys1939_r1998},
Cox~\citep{cox1946,cox1961},
Jaynes~\citep{jaynes1954,jaynes1959,jaynes1994_r2003},
Tribus~\citep{tribus1969},
de~Finetti~\citep[\sect~4]{definetti1979},
H\'ajek~\citep{hajek2003}, and apparently
Keynes~\citep{keynes1921}. 

A plausibility density for propositions $A_{x}$ concerning
(in a limit sense) a continuous parameter $x$ is denoted
by $\pf(A_{x} \cond B)$, and is as usual implicitly
defined by
\begin{equation}
  \label{eq:def_densities}
  \pr(A_{x \in \varXi} \cond B)
  =
  \int_{\varXi} \pf(A_{x} \cond B)\,\di x
  \qquad\text{for all appropriate $\varXi$};
\end{equation}
in general, $x \mapsto \pf(A_{x} \cond B)$ is a
generalised function, here always intended in the sense of
Egorov~\citep{egorov1990,egorov1990b,demidov2001} (see
also~\citep{lighthill1958_r1964,delcroixetal2002,delcroixetal2004,oberguggenberger2001}).\footnote{Alternatively,
  integrals as the above can be intended as generalised
  Riemann
  integrals~\citep{swartz2001,bartle2001,pfeffer1993}
(see
also~\citep{bartle1996,bruckner1978,mcleod1980,mawhin1981,pfeffer1986,pfeffer1987,pfeffer1988,lamoreauxetal1998}).}

The remaining notation follows \textsc{ISO}~\citep{iso1993} and
\textsc{ANSI/IEEE}~\citep{ieee1993} standards.


\chapter{General setting of the question}
\label{sec:intro}

The question that we are going to touch is, in very
general terms, that of induction: \emph{In a given
  situation, given a collection
  of observations of particular events, assign the
  plausibility for another collection of unobserved
  events.} Note that temporal distinctions are not
relevant (we did not say `past' or `future' events) and we
shall not make any. Of course, in stating this general
question we have in mind `similar' events.\footnote{But
  `similar' in which sense? The answer to this question
  cannot be given by plausibility theory, but can be
  formalised within it, as shown later.} So let us call
these events `instances of the same phenomenon', following
de~Finetti's appropriate
terminology~\citep{definetti1931,definetti1937_t1964,definetti1979}.
We also suppose to know that each such instance can
`manifest itself' in a constant (and known) number of
mutually exclusive and exhaustive `forms', and there is a
clear similarity between the forms of each instance (that
is one of the reasons we call the events `similar').

All this can be restated and made more concrete using a
terminology that is nearer to physics; but we must keep in
mind that the setting has not therefore become less
general. We call the phenomenon a
`measurement',\footnote{Even more appropriate, but too
  long, would be `measurement scheme'.} and its instances
`measurement instances'. The forms will be called
`(measurement) outcomes'. We can finally state our
question thus: \emph{In a given situation, given the
  observed outcomes of some instances of a particular
  measurement, assign the plausibility for the unobserved
  outcomes of some other instances of that measurement.}

We represent the situation, measurements, \etc\ by
propositions. The situation by $\yI$, the measurement
instances by $\yM^{(\yo)}$, the outcomes of the $\yo$th
instance by $\yR^{(\yo)}_i$ (different instances of the same
outcome are those with different $\yo$ but identical $i$).
Instances are thus generally denoted by an index ${(\yo)}$
with $\yo=1,2,\dotsc$; its range may be infinite or finite,
a detail that will be always specified as it will be very
important in later discussions. Other propositions will be
introduced and defined later. With this representation,
our question above simply becomes the assignment of the
plausibility
\begin{equation}
  \label{eq:assign}
 \pr(
\yR^{(\yo_{\yN + \yL})}_{i_{\yN + \yL}} \land \dotsb \land
\yR^{(\yo_{\yN + 1})}_{i_{\yN + 1}}
\cond
\yM^{(\yo_{\yN + \yL})} \land \dotsb \land \yM^{(\yo_{\yN + 1})}
\land
\yR^{(\yo_{\yN})}_{i_{\yN}} 
\land \dotsb \land 
\yR^{(\yo_{1})}_{i_{1}} 
\land \yI)
\end{equation}
for all possible distinct $\yo_a$, distinct $i_a$, $\yN \ge
0$, and $\yL > 0$. 

\paragraph{A more convenient notation.}
\label{sec:inter_R_M}
If you are wondering what those $\yM^{(\yo)}$ are doing in
the context of the plausibility, consider that the
plausibility of observing a particular outcome at the
$\yo$ instance of a measurement is, in general, the
plausibility of the outcome \emph{given that} the
measurement is made, times the plausibility that the
measurement is made (if the latter is not made the outcome
has nought plausibility by definition):
\begin{equation*}
  \label{eq:real_plaus_out}
  \begin{split}
    \pr(\yR \cond \yI) &= \pr(\yR \cond \yM \land \yI)\,
    \pr(\yM \cond \yI) +  \pr(\yR \cond \lnot\yM \land \yI)\,
    \pr(\lnot\yM \cond \yI),
\\
&= \pr(\yR \cond \yM \land \yI)\,
    \pr(\yM \cond \yI).
  \end{split}
\end{equation*}
But of course when we ask for the plausibility of an
outcome we implicitly mean: \emph{given that} the
corresponding measurement is or will be made. We assume
that knowledge of an outcome implies knowledge that the
corresponding measurement has been made --- symbolically,
$\pr(\yM^{(\yo')} \cond \yR^{(\yo'')} \land \yI) = 1$ if
$\yo'= \yo''$ --- hence there is no need of specify in the
context the measurements of those outcomes that are
already in the context. But the necessity remains of
explicitly writing in the context the measurements of the
outcomes outside the context. This can sometimes be
notationally very cumbersome, and therefore we introduce
the symbol $\yme$ with the following convention: $\yme$,
in the context of a plausibility, always stands for the
conjunction of
all measurement instances $\yM^{(\yo)}$ corresponding to the
outcomes on the left of the conditional symbol `$\cond$'.
Thus, \eg,
\begin{equation*}
  \label{eq:conv_M}
  \pr(\yR^{(7)}_5 \land \yR^{(3)}_1 \cond \yme \land \yR^{(2)}_8  \land \yI)
  \equiv
  \pr(\yR^{(7)}_5 \land \yR^{(3)}_1 \cond 
 \yM^{(7)} \land \yM^{(3)} \land 
\yR^{(2)}_8  \land\yI).
\end{equation*}
Note that $\yme$ has not a constant value (it is in a
sense a metavariable); we indicate this by the use of a
different typeface (Euler Fraktur).

With the new notational convention the
plausibility~\eqref{eq:assign} can be rewritten as
\begin{equation}
  \labelbis{eq:assign}
 \pr(
\yR^{(\yo_{\yN + \yL})}_{i_{\yN + \yL}} \land \dotsb \land
\yR^{(\yo_{\yN + 1})}_{i_{\yN + 1}}
\cond \yme \land 
\yR^{(\yo_{\yN})}_{i_{\yN}} \land \dotsb \land 
\yR^{(\yo_{1})}_{i_{1}} 
\land \yI).
\end{equation}

\chapter{The approach through infinite exchangeability}
\label{cha:appr_inf_exch}

The propensitor (and, roughly in the same way, the
frequentist as well) approaches the question of assigning
a value to~\eqref{eq:assign} by supposing that the outcome
instances (`trials') are `i.i.d.': that they are
independently `produced' with constant but `unknown'
propensities. As $\yN$ in~\eqref{eq:assign} becomes large,
the relative frequencies of the outcomes \emph{must} tend
to the numerical values of the propensities. These
frequencies can then be used as `estimates' of the
propensities, and so the estimated propensity for outcome
$\yR_i$ at an additional measurement instance can be
given. This summary surely appears laconic to readers that
have superficial knowledge of this practise. But this does
not matter: it is the approach of the de~Finettian that
interests us.\footnote{There are authors who apparently
  keep a foot in both camps; see \eg\ Lindley and
  Phillips' article~\citep{lindleyetal1976}} Recall that
for the de~Finettian, and for the Bayesian in general,
`probability' is not a physical concept like `pressure',
but a logical (and subjective) one like `truth'. To mark
this difference in concept we are using the term
\emph{plausibility}, which has a more logical and
subjective sound.

How would a de~Finettian approach the problem above? More
or less as follows:

\bigskip

\emph{[A fictive de~Finettian speaking:]} `Before I know of any
measurement outcomes, I imagine all possible
\emph{infinite} collections of instances of the
measurement $\yM$. Let us suppose that I judge two
collections that have the same frequencies of outcomes to
have also the same probability. The probability
distribution that I assign to the infinite collections of
outcomes is therefore symmetric with respect to exchanges
of collections having the same frequencies. Such a
distribution is called \emph{infinitely exchangeable}.
De~Finetti's representation
theorem~\citep{definetti1937_t1964,hewittetal1955,definetti1970_t1990,definetti1970b_t1990,heathetal1976,diaconis1977,diaconisetal1980,jaynes1986c,bernardoetal1994}
(see also Johnson and Zabell~\citep{johnson1924,johnson1932c,zabell1982})
says that any $\yL$-outcome marginal of such distribution, where the
outcomes $\set{\yR_i}$ appear with relative frequencies $\yLf \equiv
(\yL_i)$, can be \emph{uniquely} written in the following form:
\begin{equation}
  \label{eq:deF_thm}
  \pr(
\underset{\text{\makebox[0pt]{$\yR_1$ appears $\yL_1$ times, \etc}}}{\underbrace{\yR^{(\yo_{\yL})}_{i_{\yL}} \land \dotsb \land  \yR^{(\yo_{1})}_{i_{1}}}}
\cond \yme \land  \yI) =
  \int \Bigl(\tprod_i \yqq_i^{\yL_i}\Bigr) \, \yP(\yq \cond \yI) \,
  \di\yq,
\end{equation}
where $\yq \equiv (\yqq_i)$ are just \emph{parameters} ---
not probabilities!\ --- satisfying the same positivity and
normalisation conditions ($\yqq_i \ge 0$, $\sum_i \yqq_i =
1$) as a probability distribution, and $\yq \mapsto
\yP(\yq \cond \yI)$ is a positive and normalised
generalised function, which can be called the
\emph{generating function} of the
representation~\citep[\cf][]{jaynes1986c}. Example: I can
write the probability of a collection of three measurement
instances with two outcomes $\yR_5$ and one $\yR_8$ as
\begin{equation*}
  \pr(\yR^{(\yo_{1})}_5 \land \yR^{(\yo_{2})}_5  \land \yR^{(\yo_{3})}_8
  \cond \yme \land  \yI) = \int {\yqq_5}^2 \, \yqq_8 \, \yP(\yq \cond \yI)
  \, \di\yq.
\end{equation*}
 In particular, the probability for the
outcome $i$ in the whatever measurement instance $\yo$
can be written as
\begin{equation}
  \label{eq:plaus_next}
  \pr(\yR^{(\yo)}_i \cond \yM^{(\yo)} \land  \yI) = \int \yqq_i \, \yP(\yq \cond \yI)
  \, \di\yq.
\end{equation}
A propensitor or a frequentist would say that the
right-hand side of \eqn~\eqref{eq:deF_thm} represents the
fact that the outcome instances are ``independent''
(therefore ``their probabilities $\yqq_i$ are simply
multiplied to give the probability of their
conjunction''), ``identically distributed'' (therefore
``the probability distributions $\yqq$ is the same for all
instances''), and moreover ``their probability is
unknown'' (therefore ``the expectation integral over all
possible probability distributions $\yq$''). But
de~Finetti's theorem shows that all these mathematical
features are simply consequences of infinite
exchangeability, and we do not need any ``i.i.d.''\
terminology. 

`The generating function $\yq \mapsto \yP(\yq \cond \yI)$
is, by de~Finetti's theorem, equal to the limit
\begin{equation}
  \label{eq:P_is_lim_freq}
  \yP(\yq \cond \yI)\, \di\yq 
=
 \lim_{\yL \to \infty} 
 \pr\left(\;\parbox{0.55\columnwidth}{\prop{All possible
       collections of $\yL$ outcomes  with frequencies in
the range       $\opop{\yL\yqq_i, \yL(\yqq_i + \di\yqq_i)}$}}\;
   \bigcond \yme \land \yI \right),
\end{equation}
\ie, $\yP(\yq \cond \yI)$ is equal to the probability ---
assigned by \emph{me} --- that my imagined infinite
collection of outcomes has limiting relative frequencies
equal to $(\yqq_i)$. My exchangeable probability
assignment determines thus $\yP$ uniquely. But
de~Finetti's theorem also says the converse, \viz, any
positive and normalisable $\yP$ uniquely determines an
infinitely exchangeable probability distribution. This
allows me to specify my probability assignment by giving
the function $\yP$ instead of the more cumbersome
probability distribution for infinite collections of
outcomes.

`Let us suppose that I am now given a collection of $\yN$
measurement outcomes, and in particular their absolute
frequencies $\yNf \equiv (\yN_i)$. Given this evidence,
the probability for a collection of further $\yL$
unobserved measurement outcomes with frequencies $(\yL_i)$
is provided by the basic rules of probability theory:
\begin{multline}
  \label{eq:new_ev}
  \pr(
\underset{\text{\makebox[0pt]{$\yR_i$ appears $\yL_i$ times}}}{\underbrace{
\yR^{(\yo_{\yN + \yL})}_{i_{\yN + \yL}} \land \dotsb \land
\yR^{(\yo_{\yN + 1})}_{i_{\yN + 1}} 
}}
\cond \yme \land
\underset{\text{\makebox[0pt]{$\yR_i$ appears $\yN_i$ times}}}{\underbrace{
\yR^{(\yo_{\yN})}_{i_{\yN}} \land \dotsb \land \yR^{(\yo_{1})}_{i_{1}} 
}}
\land \yI) ={}
\\[2\jot]
\frac{
 \pr(\yR^{(\yo_{\yN + \yL})}_{i_{\yN + \yL}} \land \dotsb \land \yR^{(\yo_{\yN + 1})}_{i_{\yN + 1}} \land
\yR^{(\yo_{\yN})}_{i_{\yN}} \land \dotsb \land
\yR^{(\yo_{1})}_{i_{1}} \cond \yme \land \yI)
}{
\pr(\yR^{(\yo_{\yN})}_{i_{\yN}} \land \dotsb \land
\yR^{(\yo_{1})}_{i_{1}} \cond \yme \land \yI)
}.
\end{multline}
Using de~Finetti's representation theorem again, this
probability can also be written as
\begin{multline}
  \label{eq:deF_new_ev}
  \pr(\yR^{(\yo_{\yN + \yL})}_{i_{\yN + \yL}} \land \dotsb \land \yR^{(\yo_{\yN + 1})}_{i_{\yN + 1}} \cond \yme \land  
\yR^{(\yo_{\yN})}_{i_{\yN}} \land \dotsb \land \yR^{(\yo_{1})}_{i_{1}}
\land \yI) =
 {}\\
\int \Bigl(\tprod_i \yqq_i^{\yL_i}\Bigr) \, 
\yP(\yq \cond 
\yR^{(\yo_{\yN})}_{i_{\yN}} \land \dotsb \land \yR^{(\yo_{1})}_{i_{1}}
\land \yI)
  \, \di\yq.
\end{multline}
The function $\yq \mapsto \yP(\yq \cond
\yR^{(\yo_{\yN})}_{i_{\yN}} \land \dotsb \land
\yR^{(\yo_{1})}_{i_{1}} \land \yI)$ is different from the
previous one $\yq \mapsto \yP(\yq \cond \yI)$, but the two
can be shown~\citep[\eg,][]{bernardoetal1994} to be related
by the remarkable formula
\begin{equation}
  \label{eq:upd_Q}
\yP(\yq \cond \yR^{(\yo_{\yN})}_{i_{\yN}} \land \dotsb \land \yR^{(\yo_{1})}_{i_{1}}
\land \yI)
  =
  \frac{
    \Bigl(\tprod_i \yqq_i^{\yN_i}\Bigr) \,
    \yP(\yq \cond \yI)
  }{
    \int \Bigl(\tprod_i \yqq_i^{\yN_i}\Bigr) \, \yP(\yq \cond \yI) \,
    \di\yq
  },
\end{equation}
which is \emph{formally} identical with Bayes' theorem if we
define
\begin{gather}
  \label{eq:R_from_q}
  \yP(
\yR^{(\yo_1)}_{i_1} \land \yR^{(\yo_2)}_{i_2}
  \land \dotsb \land
\yR^{(\yo_\yL)}_{i_\yL}  
\cond \yq, \yI) \defd
\yqq_{i_1} \yqq_{i_2} \dotsm \yqq_{i_\yL} 
  \quad\text{for all $\yL$, $i_a$, and distinct
    $\yo_a$,}
\\
\intertext{which includes in particular}
  \yP(\yR^{(\yo)}_i \cond \yq,  \yI) \defd \yqq_i
  \quad\text{for all $\yo$.}
\end{gather}
(Note that these are only \emph{formal} definitions and
not probability judgements, since the various $\yP$s are
not probability distributions.) Thus I can not only
specify my infinitely exchangeable distribution by $\yP$,
but also update it by `updating' $\yP$. Moreover, for
$\yN$ enough large this function has the limit
\begin{equation}
  \label{eq:upd_Q_limit}
\yP(\yq \cond \underset{\text{\makebox[0pt]{$\yR_i$ appears $\yN_i$ times}}}{\underbrace{
\yR^{(\yo_{\yN})}_{i_{\yN}} \land \dotsb \land \yR^{(\yo_{1})}_{i_{1}} 
}} \land \yI) \simeq
\delt\Bigl(\tfrac{\yNf}{\yN} - \yq\Bigr)
\quad \text{as $\yN \to \infty$.}
\end{equation}
Comparing with \eqn~\eqref{eq:plaus_next}, this means
that
\begin{equation}
  \label{eq:plaus_next_large_N}
  \pr(\yR^{(\yo_{\yN+1})}_i \cond  \yM^{(\yo_{\yN+1})} \land 
\underset{\text{\makebox[0pt]{$\yR_i$ appears $\yN_i$ times}}}{\underbrace{
\yR^{(\yo_{\yN})}_{i_{\yN}} \land \dotsb \land \yR^{(\yo_{1})}_{i_{1}} 
}} \land\yI) 
\simeq \frac{\yN_i}{\yN}
\quad \text{as $\yN \to \infty$\quad(with $\yo \ne \yo_a$),}
\end{equation}
\ie, the probability I assign to an unobserved outcome
gets very near to its observed relative frequency, as
observations accumulate. This also means that two persons
having different but compatible initial beliefs (\ie,
different initial exchangeable distributions having the
same support) and sharing the same data, tend to converge
to similar probability assignments.'

\bigskip

The point of view summarised above by the de~Finettian is
quite powerful. It allows the de~Finettian to make sense,
without the need of bringing along ugly or meaningless
metaphysical concepts, of those mathematical expressions
very often used by propensitors (and frequentists as well)
that are formally identical to
formulae~\eqref{eq:plaus_next} (`expected propensity' or
`estimation of unknown probability'),
\eqref{eq:P_is_lim_freq} (`probability as limit
frequency'), \eqref{eq:R_from_q} (definition of
`propensity'), \eqref{eq:plaus_next_large_N} (`probability
equal to past frequency'). This point of view and the
notion of exchangeability can moreover be generalised to
more complex
situations~\citep{hewittetal1955,heathetal1976,diaconis1977,definetti1938_t1980,diaconisetal1980,jaynes1986c,diaconisetal1987,georgii1979}\citep[\cf\
also][]{ladetal1990}, leading to other powerful
mathematical expressions and techniques.

\chapter{Why a complementary approach?}
\label{sec:why_appr}

We shall presently present and discuss another approach,
not based on exchangeability or representation theorems,
that can be used to give an answer to the question of
induction, and to make sense of the propensitors' and
frequentists' formulae and practise as well.
Why did we seek an approach different from that based on
infinite exchangeability? Here are some reasons:

\begin{asparaenum}[(a)]
  \item\label{item:uncertain}One sometimes feels
  `uncertain', so to speak, about one's plausibility
  assignment. One can also say that some plausibility
  assignments feel sometimes more `stable' than others.
  This is lively exemplified by
  Jaynes~\citep[\chap~18]{jaynes1994_r2003}:
  \begin{quotation}
    Suppose you have a penny and you are allowed to
    examine it carefully, convince yourself that it's an
    honest coin; \ie\ accurately round, with head and
    tail, and a center of gravity where it ought to be.
    Then, you're asked to assign a probability that this
    coin will come up heads on the first toss. I'm sure
    you'll say $1/2$. Now, suppose you are asked to assign
    a probability to the proposition that there was once
    life on Mars. Well, I don't know what your opinion is
    there, but on the basis of all the things that I have
    read on the subject, I would again say about $1/2$ for
    the probability. But, even though I have assigned the
    same ``external'' probabilities to them, I have a very
    different ``internal'' state of knowledge about those
    propositions.

    To see this, imagine the effect of getting new
    information. Suppose we tossed the coin five times and
    it comes up tails every time. You ask me what's my
    probability for heads on the next throw; I'll still
    say $1/2$. But if you tell me one more fact about Mars,
    I'm ready to change my probability assignment
    completely. There is something which makes my state of
    belief very stable in the case of the penny, but very
    unstable in the case of Mars.
  \end{quotation}
  An example similar to the Martian one is that of a trickster's coin,
  which always comes up heads or always tails (either because the coin is
  two-headed or two-tailed, or because of the tosser's skills). Not knowing
  which way the tosses are biased, you assign $1/2$ that the coin will come
  up heads on the first observed toss. But as soon as you see the outcome,
  your plausibility assignment for the next toss will
  collapse\footnote{Just like a quantum-mechanical wave-function.} to $1$
  or $0$. The `uncertainty' in the plausibility assignments given to the
  penny and to the trickster coin's toss could be \emph{qualitatively}
  pictured as in Fig.~\ref{fig:coins}. We think that this `uncertainty' in
  the plausibility assignment --- or better, as Jaynes calls it, this
  `difference in the internal state of knowledge' with regard to the
  propositions involved, is a familiar and undeniable feeling. There is in
  fact a rich literature which tries to take it into account by exploring
  or even proposing alternative plausibility theories based on probability
  intervals or probabilities of probabilities (see
  \eg~\citetext{\citealp{atkinsonetal1964,jamison1970,levi1974,levi1984,kyburg1987};
    \citealp{ladetal1990}, esp.\ \sect~3.1; \citealp{fishburn1983,nau1992}}
  and \cf~\citep[\sect~2.2]{good1965}).
\begin{figure}[bt!]
\includegraphics[width=\columnwidth]{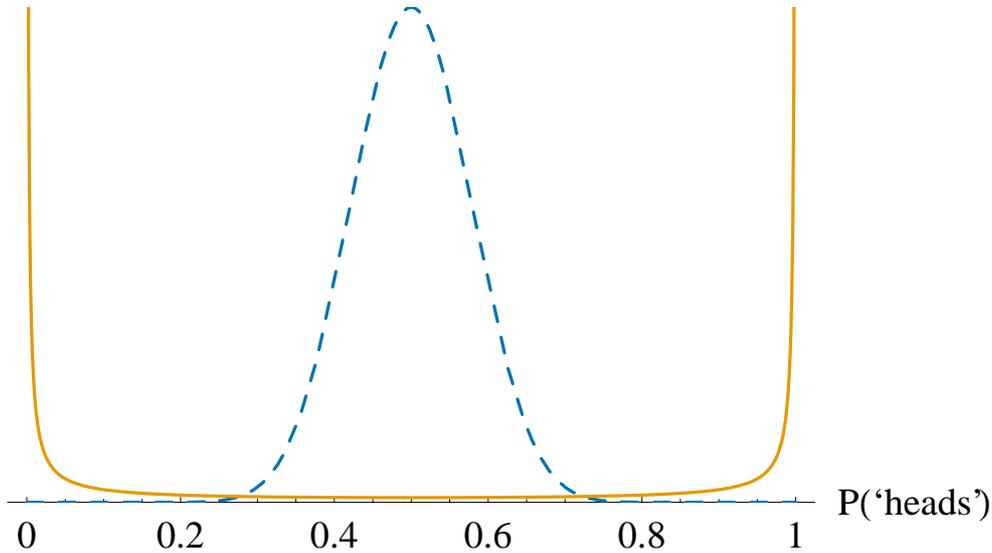}
\caption{\label{fig:coins}Qualitative illustration of the
  `uncertainty' in the plausibility assignments for
  `heads' in the case of a fair coin-toss (blue dashed
  curve, with maximum at $1/2$) and the toss of a
  trickster's coin (orange continuous curve, with maxima
  at $0$ and $1$). The illustration is made quantitative
  in \sect~\ref{sec:sureness}.}
\end{figure}

\item\label{item:motiv_exch}A plausibility assignment,
according to de~Finetti's approach, begins with a
judgement of exchangeability. One is not concerned
\emph{within the formalism itself} with the motivation of
such a judgement (de~Finetti recognises that there usually
is a motivation, but it is relegated to the informal
meta-theoretical considerations). Yet such judgements are
often grounded, especially in natural philosophy, on very
important\footnote{And, trivially, subjective.} reasonings
and motivations which one would like to analyse by means
of plausibility theory.\footnote{In a way there is a point
  in Good's statement~\citep[\chap~3]{good1965} `it seems
  to me that one would not accept the [exchangeability
  assumption] unless one already had the notion of
  physical probability and (approximate) statistical
  independence at the back of one's mind', although there
  is no need to bring `physical probabilities' and
  `statistical independence' along.}


\item\label{item:finite_case}When the maximum number of
possible observations (measurements) is finite \emph{and
  small} it does not make sense to make a judgement of
infinite exchangeability, not even as an approximation
(\cf~\citep{diaconis1977}\citep[\sect~4.7.1]{bernardoetal1994}).
De~Finettians can in this case use \emph{finite}
exchangeability~\citep{diaconis1977,diaconisetal1980}, but
then they cannot make sense of the propensitor's formulae
like~\eqref{eq:plaus_next} or~\eqref{eq:R_from_q}, which
the propensitor, however, is still entitled to use even in
this case.

\item\label{item:quantum}Finally, another important
reason is that the generalisation of de~Finetti's theorem
studied by Caves, Fuchs, and
Schack~\citep{cavesetal2002,cavesetal2002b,fuchsetal2004,fuchsetal2004b}
fails for some physical statistical models, \viz\ quantum
mechanics on real and quaternionic Hilbert space. We
consider this failure not as a sign that complex quantum
theory is somehow `blessed', but rather as a sign that
either the theorem can be generalised in some other way
that holds in any physical statistical model, or
de~Finetti's approach can be substituted or complemented
by another one whose generalisation applies to any
physical statistical model whatever. The point here is
that de~Finetti's theorem --- just like the whole of
plausibility theory --- belongs to the realm of
\emph{logic}, not physics, and thus should apply to any
conceivable physical theory that be logically consistent,
even one that does not describe actual phenomena.
\end{asparaenum}

  The alternative point of view to be presented meets all
  the points raised above:
  \begin{inparaenum}[(a)]
  \item It allows us to formalise within plausibility
    theory --- \ie, without resorting to interval-valued
    plausibilities and the like --- the intuitive notion
    of `uncertainty' of a plausibility assignment, and
    even to quantify it. \item It stems directly from an
    analysis of the conditions in which judgements of
    exchangeability usually originate and
    apply. 
    \item It can be used to interpret the propensitor's
    practise when the maximum number of possible
    observations is finite and small. \item It readily
    generalises to any logically consistent physical
    statistical model, whether it apply to actual
    phenomena or not; in particular, within quantum theory
    it easily applies to such problems as quantum-state
    assignment, quantum-state `teleportation', and others
    that involve the notion of `unknown quantum state'.
  \end{inparaenum}

  This point of view, which we shall call the `Laplace-Jaynes approach',
  stems from a re-reading and possibly a re-interpretation of
  Jaynes~\citep[esp.\ pp.~11, 12, 15]{jaynes1986d_r1996} (\cf\
  also~\citep[lect.~18]{jaynes1954}\citep[lect.~5]{jaynes1959}\citep[\chap~18]{jaynes1994_r2003}),
  Laplace~\citep{laplace1774}, and
  de~Finetti~\citep[\sect~20]{definetti1931} (\cf\ also
  Caves~\citep{caves2000c}) and is strictly related to the `approach
  through circumstances' presented in the previous
  paper~\citep{portamanaetal2006}. Mosleh and Bier~\citep{moslehetal1996}
  also propose and discuss basically the same point of view.

  It cannot be too strongly emphasised that this point of
  view is fully in line with plausibility theory, Bayesian
  theory, and de~Finetti's point of view: Some
  plausibilities with particular properties will be
  introduced, based on particular judgements; but the
  acceptance or not of the latter is, as with judgements
  of exchangeability, always up to the individuals and
  their knowledge.

  We now present this `Laplace-Jaynes approach' and show
  how it can be used for the question of induction in a
  manner parallel to the approach through infinite
  exchangeability. We shall then discuss how it meets
  points~\ref{item:uncertain}, \ref{item:motiv_exch},
  \ref{item:finite_case} above (not necessarily in that
  order). The discussion of the fourth point is left to
  the next study of this series (in preparation).


\chapter{The Laplace-Jaynes approach}
\label{sec:exch_wh}

\section{Introducing a set of circumstances}
\label{sec:intro_circ}

With the notation and the general settings of the
introduction, let us recapitulate how we reason and
proceed in the case of exchangeability. The
situation 
$\yI$ leads us to assign an infinitely exchangeable
plausibility distribution to the collection of measurement
outcomes. In other words, $\yI$ is such that we consider
two plausibilities like \eg\ $\pr(\yR^{(2)}_5 \land
\yR^{(4)}_7 \cond \yme \land  \yI)$ and $\pr(\yR^{(4)}_5 \land
\yR^{(2)}_7 \cond \yme \land  \yI)$ as equal.

In \sect~\ref{sec:why_appr}, point~\eqref{item:motiv_exch},
we remarked that the approach through exchangeability is
not formally concerned about those details of the
situation which lead us to see a sort of analogy ar
similarity amongst different measurement and outcome
instances and thence make a judgement of exchangeability.
Rather, the whole point is that this similarity is
expressed by, or reflected in, the exchangeable plausibility
assignment. As de~Finetti says, `Our reasoning will only
bring in the events, that is to say, the trials, each
taken individually; the analogy of the events does not
enter into the chain of reasoning in its own right but
only to the degree and in the sense that it can influence
in some way the judgment of an individual on the
probabilities in
question'~\citep[p.~120]{definetti1937_t1964}. And also:
\begin{quote}
  What are sometimes called \emph{repetitions} or
  \emph{trials}\footnote{And what we call here
    \emph{instances} (Authors' Note).} of the same event
  are for us distinct events. They have, in general, some
  common characteristics or symmetries which make it
  natural to attribute to them equal probabilities, but we
  do not admit any \emph{a priori} reason which prevents
  us in principle from attributing to each of these trials
  [\ldots]\
  some different and absolutely arbitrary probabilities
  [\ldots].
  In principle there is no difference for us between this
  case and the case of $n$ events which are not analogous to
  each other; the analogy which suggests the name ``trials
  of the same event'' (we would say ``of the same
  phenomenon'') is not at all essential, but, at the most,
  valuable because of the influence it can exert on our
  psychological judgment in the sense of making us
  attribute equal or very nearly equal probabilities to
  the different events.'~\citep[p.~113,
  footnote~1]{definetti1937_t1964}
\end{quote}

But although we agree on this terminology and its
motivation, we do not agree on the unqualified and
indiscriminate diminution of the importance of the
similarity amongst measurement instances. Such similarity,
especially in the natural sciences, often stems from or is
traced back to similarities at deeper\footnote{We do not
  mean `microscopic'.} levels of analysis and
observation. It is by this process that natural philosophy
proceeds.

So let us suppose that the considerations that lead us to
a probability assignment for various measurement instances
can be analysed --- and formalised --- at a deeper
level.\footnote{Using, once more, words by de~Finetti: `we
  enlarge the analysis to include our state of mind in
  relation to other events, from which it might or might
  not be \emph{independent} and by which, consequently, it
  will or will not be modified if they occur, or if we
  learn of their
  occurrence'~\citep[\sect~28]{definetti1931_t1989}.} More
precisely, we suppose to have identified for each
measurement instance $\yo$ a set of diverse possible
`circumstances' $\set{\yC^{(\yo)}_j \with j = 1, \dotsc}$
--- all these sets having the same cardinality --- with
the following properties:

\begin{enumerate}[I.]
\item The first involves the plausibilities we assign to
  the circumstances:
  \begin{gather}\label{eq:exc_exh}\tag{I}
    \begin{aligned}
      &\pr(\yC^{(\yo)}_{j} \cond \yD \land \yI) =
      \pr(\yC^{(\yo)}_{j} \cond \yI)\quad
      \parbox[t]{0.5\columnwidth}{for all $j$ and all
        $\yD$ representing conjunctions of
        measurements 
        (but \emph{not} of outcomes!) from the same or
        other instances,}
      \\[1.5\jot]
      &\pr(\yC^{(\yo)}_{j'} \land \yC^{(\yo)}_{j''} \cond \yI)
      = 0 \quad\text{if $j' \neq j''$},
      \\[1.5\jot]
      &\pr\bigl(\tlor_j \yC^{(\yo)}_{j} \cond \yI \bigr) =
      \tsum_j \pr(\yC^{(\yo)}_{j} \cond \yI) = 1,
    \end{aligned}
  \end{gather}
  \ie, we judge the $\set{\yC^{(\yo)}_j \with j =
    1, \dotsc}$ to be mutually exclusive and exhaustive, or
  in other words we are certain that one of them holds,
  but we do not know which. Moreover, we judge any
  knowledge of measurement instances (but not of outcome
  instances) as irrelevant for assessing the
  plausibilities of these circumstances, or in other
  words, simply knowing that a measurement is made does
  not give us clues as to the exact circumstance in which
  it is made.\footnote{This assumption can be relaxed.}

 \item
  The second involves the plausibility distribution we
  assign to the outcomes, conditional on knowledge of a
  circumstance:
  \begin{multline}
    \label{eq:screening}\tag{II}
    \pr(\yR^{(\yo)}_{i} \cond \yM^{(\yo)} \land \yD \land
    \yC^{(\yo)}_{j} \land \yI) = \pr(\yR^{(\yo)}_{i} \cond
    \yM^{(\yo)} \land \yC^{(\yo)}_{j} \land \yI) 
    \\
    \parbox[b]{0.8\columnwidth}{for all $\yo$, $i$, $j$, and
      all $\yD$ representing a conjunction of
      measurements, measurement outcomes, and
      circumstances of instances different from $\yo$,}
  \end{multline}
  which means that if we were certain about any
  circumstance $\yC^{(\yo)}_j$ for the instance $\yo$ we
  should judge knowledge of any data concerning other
  instances as irrelevant for the assessment of the
  plausibility of the outcome of instance $\yo$. (The
  expression above is undefined if $\yC^{(\yo)}_j$ happens
  to be inconsistent with $\yD$, \ie\ if $\pr(\yD \cond
  \yme \land \yC^{(\yo)}_j \land \yI) =0$; but this will not
  cause problems in the following analysis.)

\item The third property concerns the relationships amongst
  our plausibility assignments for the outcomes in
  different instances:
  \begin{equation}
    \label{eq:ind_inst}\tag{III}
    \pr(\yR^{(\yo')}_{i} \cond \yM^{(\yo')} \land \yC^{(\yo')}_{j} \land
    \yI) 
    =
    \pr(\yR^{(\yo'')}_{i} \cond \yM^{(\yo'')} \land \yC^{(\yo'')}_{j} \land
    \yI)
    \defs \yqq_{ij}
\quad\text{for all $\yo'$, $\yo''$,} 
\end{equation}
  which expresses the fact that we see a similarity
  between outcomes and circumstances of different
  instances, like when we say `the same outcome', `the
  same measurement', or `the same circumstance'. It is
  thanks to this property that we can often drop the
  instance index `$(\yo)$' and make sense of expressions
  like `$\pr(\yR_{i} \cond \yM \land \yC_{j} \land \yI)$',
  which can stand generically for
  \begin{equation}
    \label{eq:no_inst_ind}
    \begin{split}
      \pr(\yR_{i} \cond \yM \land \yC_{j} \land \yI)
      &\defd \pr(\yR^{(\yo)}_{i} \cond \yM^{(\yo)} \land
      \yC^{(\yo)}_{j} \land \yI) \quad \text{for any $\yo$,}
      \\
      &\equiv \yqq_{ij}.
    \end{split}
  \end{equation}
  In the following, expressions like `$\pr(\yR_{i} \cond
  \yM \land \yC_{j} \land \yI)$' will be understood in the
  above sense.

\item The fourth property strengthen the similarity
  amongst instances, and is the one that makes induction
  possible:
  \begin{equation}
    \label{eq:induction}\tag{IV}
    \pr(\yC^{(\yo')}_{j'} \cond \yC^{(\yo'')}_{j''} \land \yI)
    =
    \delt_{j'\,j''} \quad \text{for all $\yo',\yo''$.}
  \end{equation}
  This means that we believe that if a particular
  circumstance holds in a particular instance, then the
  `same' circumstance (in the sense of~\eqref{eq:ind_inst}
  and~\eqref{eq:induction}) holds in all other instances.
  This property implies (and, with the help
  of~\eqref{eq:exc_exh}, is implied by) the following:
  \begin{equation}
    \label{eq:def_join_j_single_j}
    \text{for all $\yo$ and $j$,}
    \quad
    \pr(\yC^{(\yo)}_{j} \cond \yI)
    =
    \pr(\tland_{\yj} \yC^{(\yj)}_{j} \cond \yI) \defs \yQ_j
  \end{equation}
  for some $\yQ_j$; and
  \begin{equation}
    \label{eq:pl_seq_circ}
    \pr(\tland_\yj \yC^{(\yj)}_{j_\yj} \cond \yI) 
    =
    \begin{cases}
      \yQ_j 
      &\text{if $j_\yj = j$ for all $\yj$,}
      \\
      0 &\text{otherwise,}
    \end{cases}
  \end{equation}
  \ie, the plausibility that we assign to the $\yo$th
  instance of the $j$th circumstance is equal to that
  assigned to the collection consisting exclusively of
  `repetitions' of the $j$th circumstance. Collections
  consisting of non-corresponding circumstances have
  nought plausibility. This allows us to define
  \begin{equation}
    \label{eq:def_pl_circ_j}
    \begin{split}
      \pr(\yC_{j} \cond \yI) &\defd
      \pr({\textstyle\Land}_\yj \yC^{(\yj)}_{j} \cond \yI),
      \\
      &\equiv \pr(\yC^{(\yo)}_{j} \cond \yI)\quad\text{for
        any $\yo$},
      \\
      &\equiv \yQ_j,
    \end{split}
  \end{equation}
  and
  \begin{equation}
    \yC_{j} \defd \tland_\yj \yC^{(\yj)}_{j}.
    \label{eq:def_Ck_as_join}
  \end{equation}
  Thanks to the definitions above the expression
  `$\pr(\yC_{j} \cond \yI)$' can be used to unambiguously
  denote the plausibility that `the ``same'' circumstance
  `$\yC_j$' holds in all measurement instances'.
\end{enumerate}
\emph{Note that in all the properties above the number of
  instances (\ie, the range of~$\yo$) can be either infinite
  or finite.}

Before further commenting the above properties, let us try
to partially answer the question: what are these
`circumstances'? The most general and precise answer is:
they are whatever (propositions concerning) facts you like
that make you assign plausibilities satisfying
properties~\eqref{eq:exc_exh}--\eqref{eq:induction}. No
more than this would really need be said. But we can
emphatically add that
the circumstances need not concern `mechanisms', `causes',
`microscopic conditions', or the like; and that they do
not\footnote{Cannot, in a logical sense; \cf\ Remark~2
  in~\citep{portamanaetal2006}.} concern 
`unknown probabilities'.\footnote{But the circumstances
  \emph{may} concern `propensities', if the latter are
  intended as sorts of (ugly) physical concepts. This
  possibility is due to the generality of plausibility
  theory which, like classical logic, does not forbid you
  to bring along and reason about unreal or even
  preposterous concepts and entities, like
  `nagas'~\citep{borgesetal1957_r1967_t2002},
  `valier'~\citep{tolkien1977_r1979}, `propensities', and
  `wave-particles', provided you do it in a
  self-consistent way.} They concern details of the
context that are unknown, but that (we judge) would have a
great weight in our plausibility assignment for the
$\set{\yR^{(\yo)}_i}$ if we only knew them; so great a
weight as to render (practically) unimportant any
knowledge of the details of other measurement instances.
As for the choice of such details, we have complete
freedom. A trivial example: the tosses of a coin which we
have not examined, but which we know for sure to be
two-headed or two-tailed. In this case the propositions
\prop{Coin is two-headed (at toss $\yo$)} and \prop{Coin is
  two-tailed (at toss $\yo$)} form, for each $\yo$, a set of
circumstances 
that satisfy property~\eqref{eq:exc_exh}. But we also know
that the coin is \emph{the same} at all tosses, which
implies property~\eqref{eq:induction}; hence we can omit
the specification `(at toss $\yo$)' without confusion. If we
knew that the coin was two-headed (at toss $\yo$) we should
assign unit plausibility to heads for all future or past
tosses, \ie,
\begin{equation*}
  \pr(\prop{Heads at toss $\yo$} \cond \yme \land  \prop{Coin
    is two-headed (at toss $\yo$)} \land \yI) = 1 
\quad \text{for all $\yo$,}
\end{equation*}
and we should consider knowledge of the outcomes of other
tosses as irrelevant.\footnote{Because we can but expect
  all tosses to give heads. Note that if a toss has given
  or will give tails, this signals a \emph{contradiction}
  in our knowledge. \Ie, some of the data we have (about
  the coin or about toss outcomes) have to be mendacious.
  But this is a problem that does not concern plausibility
  theory. Like logic, it can give sensible answers only if
  the premises we put in are not inconsistent.} An
analogous discussion holds for the two-tails possibility.
Thus properties~\eqref{eq:screening}
and~\eqref{eq:ind_inst} are also satisfied.

This was an extreme example, for the plausibilities
conditional on the circumstances were nought or one. But
it needs not be so: other circumstances could lead to less
extreme conditional plausibility judgements. In general,
remaining within the coin toss example, we could ask: By
which method is the coin tossed? What are its physical
characteristics (two-headedness, centre-of-mass position,
elastic and rigid properties, \etc)? Upon what is it
tossed? Who tosses it?\footnote{See Jaynes' insightful and
  entertaining
  discussion~\citep[\chap~10]{jaynes1994_r2003} on these
  kinds of factors.} 
Can there be any symmetries in my state of knowledge in
respect of the situation? What are the consequences of the
toss or of the outcomes? --- and a set of circumstances
could be distilled from the possible answers to these and
other questions. Note in particular, with regard to the
last two questions, that a circumstance may be a judgement
of symmetry or may also be a \emph{consequence}, in some
sense, of the outcomes. Such kinds of circumstances are
perfectly fine as long as they lead you to make a
plausibility assignment on the outcomes and satisfy the
properties~\eqref{eq:exc_exh}--\eqref{eq:induction}. This
emphasises again the fact that a circumstance needs not be
a sort of `mechanism' or `cause' of the measurements or of
the outcomes.
Note also that properties~\eqref{eq:exc_exh}
and~\eqref{eq:induction} determine neither the
plausibilities $\pr(\yC^{(\yo)}_j \cond \yI)$ nor the
$\pr(\yR^{(\yo)}_i \cond \yM^{(\yo)} \land \yC^{(\yo)}_j \land
\yI)$. These plausibilities are, a de~Finettian would say,
`fully subjective'.

By properties~\eqref{eq:screening} and~\eqref{eq:ind_inst}
we can interpret and give a meaning to the locution
`independent and identically distributed events'. Such
locution means only that we are entertaining some
circumstances that, according to our judgement, render the
conditional plausibilities of corresponding outcomes of
different measurement instances equal, so that we need not
specify the particular instance (`identically
distributed'); 
and render knowledge of outcomes of other instances
irrelevant, so that we can specify the conditional
plausibility for an outcome independently of the knowledge
of other outcomes (`independent'). The locutions `unknown
probability' and `probability of a probability' will also
be interpreted in a moment (\sect~\ref{sec:coarse-gr}).

\section{Approaching the problem of induction through a set of circumstances}
\label{sec:iduct_by_circ}

How do we face the question of induction with these
`circumstances' and the assumptions that accompany them?
We answer this question in two steps. 

Having introduced sets of circumstances $\set{\yC^{(\yo)}_j}$,
we must assign (subjectively, a de~Finettian would say)
for all $i$ and $j$ the plausibilities
 \begin{align}
   \label{eq:ex_Rassi}
&   \pr(\yR_i \cond \yM \land \yC_j \land \yI) \equiv
   \yqq_{ij}
&&\text{($\defd \pr(\yR^{(\yo)}_{i} \cond \yM^{(\yo)} \land \yC^{(\yo)}_{j}
    \land \yI)$ for any $\yo$),}
   \\
\label{eq:ex_Cassi}
&\pr(\yC_j \cond \yI) \equiv \yQ_j
&&\text{($\defd \pr({\textstyle\Land}_\yj \yC^{(\yj)}_{j} \cond \yI)
\equiv \pr(\yC^{(\yo)}_{j} \cond \yI)$ for any $\yo$),}
\end{align}
in the sense of \eqns~\eqref{eq:no_inst_ind}
and~\eqref{eq:def_pl_circ_j}. It is then a simple
consequence of the rules of plausibility theory, together
with the properties and definitions of the previous
sections, that the plausibility we assign to any
collection of measurement outcomes is given by
\begin{multline}
  \label{eq:decomp_j}
  \pr(
\underset{\text{\makebox[0pt]{$\yR_1$ appears $\yL_1$ times, \etc}}}{\underbrace{\yR^{(\yo_{\yL})}_{i_{\yL}} \land \dotsb \land  \yR^{(\yo_{1})}_{i_{1}}}}
\cond \yme \land  \yI) =
{}\\
\sum_j \Bigl[\tprod_i  \pr(\yR_i \cond \yM \land  \yC_j \land
\yI)^{\yL_i} \Bigr] \, \pr(\yC_j \cond \yI)
\equiv
\sum_j \Bigl(\tprod_i \yqq_{ij}^{\yL_i}\Bigr) \, \pr(\yC_j \cond \yI).
\end{multline}
Note how this plausibility assignment depends only on the
frequencies $(\yL_i)$ of the outcomes: it is an
(infinitely) exchangeable assignment --- although
exchangeability was not our starting assumption.

Let us suppose that we are now given a collection of $\yN$
measurement outcomes, and in particular their absolute
frequencies $\yNf \equiv (\yN_i)$. Given this evidence,
the plausibility for a collection of outcomes, with
frequencies $(\yL_i)$, of further $\yL$ measurements is
also derived by the basic rules of plausibility theory,
from our initial plausibility
assignments~\eqref{eq:ex_Rassi} and \eqref{eq:ex_Cassi},
using expression~\eqref{eq:decomp_j} and the properties of
the circumstances:
\begin{multline}
  \label{eq:new_ev_j}
  \pr(
\underset{\text{\makebox[0pt]{$\yR_i$ appears $\yL_i$ times}}}{\underbrace{
\yR^{(\yo_{\yN + \yL})}_{i_{\yN + \yL}} \land \dotsb \land
\yR^{(\yo_{\yN + 1})}_{i_{\yN + 1}} 
}}
\cond \yme \land
\underset{\text{\makebox[0pt]{$\yR_i$ appears $\yN_i$ times}}}{\underbrace{
\yR^{(\yo_{\yN})}_{i_{\yN}} \land \dotsb \land \yR^{(\yo_{1})}_{i_{1}} 
}}
\land \yI) 
={}
\\[2\jot]
\sum_j \Bigl[\tprod_i  \pr(\yR_i \cond \yM \land  \yC_j \land
\yI)^{\yL_i} \Bigr] \, \pr(\yC_j \cond 
\yR^{(\yo_{\yN})}_{i_{\yN}} \land \dotsb \land \yR^{(\yo_{1})}_{i_{1}}
\land
\yI)
\equiv{}
\\
\sum_j \Bigl(\tprod_i \yqq_{ij}^{\yL_i}\Bigr) \, \pr(\yC_j
\cond 
\yR^{(\yo_{\yN})}_{i_{\yN}} \land \dotsb \land \yR^{(\yo_{1})}_{i_{1}}
\land
\yI),
\end{multline}
with
\begin{equation}
  \label{eq:upd_pr_Cj}
\pr(\yC_j \cond \yR^{(\yo_{\yN})}_{i_{\yN}} \land \dotsb \land \yR^{(\yo_{1})}_{i_{1}}
\land\yI)
  =
  \frac{
    \Bigl(\tprod_i \yqq_{ij}^{\yN_i}\Bigr) \,
    \pr(\yC_j \cond \yI)
  }{
    \tsum_j \Bigl(\tprod_i \yqq_{ij}^{\yN_i}\Bigr) \, 
\pr(\yC_j \cond \yI) 
  }.
\end{equation}

These formulae present many similarities to the
de~Finettian's~\eqref{eq:deF_new_ev} and~\eqref{eq:upd_Q},
and to the formally similar expressions used by
`frequentists' and `propensitors'. But it should be noted
that, in the present formulae, $\yqq_{ki} \defd \pr(\yR_i
\cond \yme \land \yC_j \land \yI)$, $\pr(\yC_j \cond
\yI)$, and $\pr(\yC_j \cond \yR^{(\yo_{\yN})}_{i_{\yN}}
\land \dotsb \land \yR^{(\yo_{1})}_{i_{1}} \land\yI)$ are
\emph{actual plausibilities}, not just parameters or
positive and normalised generating functions. (`And they
are fully subjective!', the de~Finettian reiterates).

\section{Plausibility-indexing the set of circumstances}
\label{sec:coarse-gr}

The marked similarities can in fact be made into
identities of form. To achieve this, we go back to the
point where, after having introduced the circumstances
$\set{\yC^{(\yo)}_j}$, we made the plausibility assignments
 \begin{align}
   \labelbis{eq:ex_Rassi}
&   \pr(\yR_i \cond \yM \land \yC_j \land \yI) \equiv
   \yqq_{ij}
&&\text{($\defd \pr(\yR^{(\yo)}_{i} \cond \yM^{(\yo)} \land \yC^{(\yo)}_{j}
    \land \yI)$ for any $\yo$),}
   \\
\labelbis{eq:ex_Cassi}
&\pr(\yC_j \cond \yI) \equiv \yQ_j
&&\text{($\defd \pr({\textstyle\Land}_\yj \yC^{(\yj)}_{j} \cond \yI)
\equiv \pr(\yC^{(\yo)}_{j} \cond \yI)$ for any $\yo$),}
\end{align}
for all $i,j$ (and $\yo$). Instead of proceeding as we did,
we now follow the remark by
Jaynes~\citep[p.~12]{jaynes1986d_r1996}, cited in the
introductory epigraph:
\begin{quote}
  It is simply convenient to index our hypotheses by
  parameters [$\yq$] chosen to be numerically equal to the
  probabilities assigned by those hypotheses; this avoids
  a doubling of our notation. We could easily restate
  everything so that the misconception could not arise; it
  would only be rather clumsy notationally and tedious
  verbally.
\end{quote}
What Jaynes calls `hypotheses' we have here interpreted, more
generically, as `circumstances'. Let us now group together
those that lead to the same plausibility distribution for
the outcomes. That is, fix a $\yo$, and form the equivalence
classes of the equivalence relation
\begin{equation}
  \label{eq:equiv_rel}
  \yC^{(\yo)}_{j'} \sim \yC^{(\yo)}_{j''} \iff
\text{for all $i$, }
\pr(\yR^{(\yo)}_i \cond \yM^{(\yo)} \land \yC^{(\yo)}_{j'} \land \yI) = 
\pr(\yR^{(\yo)}_i \cond \yM^{(\yo)} \land \yC^{(\yo)}_{j''} \land \yI).
\end{equation}
By the very method these classes are defined, each one can
be \emph{uniquely} identified by a particular set of
values $(\yqq_i)\equiv \yq$ of the plausibility
distribution for the outcomes. Note that owing to
property~\eqref{eq:ind_inst} the value of $\yq$ does not
depend on $\yo$. Call therefore $\yqc$ the class identified
by a particular $\yq$, and denote membership of $\yC^{(\yo)}_j$ by
`$j \in \yqc$' for short.
Now let us take the disjunction of the circumstances in
each class $\yqc$ and uniquely denote this disjunction by
$\yS^{(\yo)}_{\yq}$:
\begin{equation}
  \label{eq:def_disj_q}
  \yS^{(\yo)}_{\yq} \defd \tlor_{j \in \yqc} \yC^{(\yo)}_j.
\end{equation}
It is easy to see that, owing again to
property~\eqref{eq:exc_exh}, each $\yS^{(\yo)}_{\yq}$ yields a
conditional distribution for the outcomes that is
numerically equal to $\yq$, \ie\ numerically equal to all
those yielded by the $\yC^{(\yo)}_j$ in
$\yqc$: 
\begin{equation}
  \label{eq:cond_plaus_Cq}
  \pr(\yR^{(\yo)}_i \cond \yM^{(\yo)} \land \yS^{(\yo)}_{\yq}
  \land \yI) = \yqq_i
  \quad\text{for any $\yo$}.
\end{equation}
Moreover, the $\set{\yS^{(\yo)}_{\yq}}$ also satisfy
properties~\eqref{eq:exc_exh}--\eqref{eq:induction} as
the circumstances $\set{\yC^{(\yo)}_j}$, and their
plausibilities are easily obtained from those of the
latter:
\begin{align}
  \label{eq:plaus_Cq_from_Cj}
  \pr(\yS^{(\yo)}_{\yq} \cond \yI) &= 
\tsum_{j \in \yqc} \pr(\yC^{(\yo)}_j \cond \yI),
&
 \pr(\yS_{\yq} \cond \yI) &= 
\tsum_{j \in \yqc} \pr(\yC_j \cond \yI),
\end{align}
where $\yS_{\yq} \defd \Land_\yj \yS^{(\yj)}_{\yq}$ in
analogy with the definition~\eqref{eq:def_Ck_as_join}.

These $\yS^{(\yo)}_{\yq}$ are thus a sort of `plausibility-indexed'
circumstances. We shall call them also `coarse-grained circumstances'
sometimes. The $\yC^{(\yo)}_{j}$ can then be called `fine-grained'
circumstances when a distinction is necessary.

In general, we have a class $\yqc$ and a disjunction
$\yS^{(\yo)}_{\yq}$ for particular (vector) values of $\yq$
only (depending on the initial choice of circumstances and
on the initial assignments~\eqref{eq:ex_Rassi}). But we
can formally introduce propositions $\yS^{(\yo)}_{\yq'}$,
all equal to the false proposition ($A \land \lnot A$),
for the remaining values $\yq'$. These propositions hence
have nought plausibilities, $\pr(\yS^{(\yo)}_{\yq'} \cond
\yI) \equiv 0$. The plausibilities $\pr(\yR^{(\yo)}_i \cond
\yM^{(\yo)} \land \yS^{(\yo)}_{\yq'} \land \yI)$ are
undefined, but they will appear multiplied by the former
in all relevant formulae, and hence their product will
vanish by convention. With this expedient we can consider
the whole, continuous set $\set{\yS^{(\yo)}_{\yq} \st \yq
  \in \ysim}$ for all possible distributions $\yq$ which
naturally belong to the simplex $\ysim \defd \set
{(\yqq_i) \st \yqq_i \ge 0, \sum_i \yqq_i =1}$. We can
thus substitute an integration $\smallint_\ysim \dotso
\pf(\yS_{\yq} \cond \dotso)\, \di\yq$ for the summation
$\sum_{\yq} \dotso \pr(\yS_{\yq} \cond \dotso)$. The
density $\pf(\yS_{\yq} \cond \dotso)$ will be a
generalised function.\footnote{We can avoid the expedient
  above if we like; then the integrals that follow must be
  understood in a measure-theoretic
  sense~\citep{kolmogorovetal1960_r1962,rudin1970,swartz1994,fremlin2000_r2004,fremlin2001_r2003,fremlin2002_r2004,fremlin2003_r2006,fremlin2003b_r2006}
  (\cf\
  also~\citep{mcshane1973,bruckner1978,zakrzewski2001}),
  with `$\pf(\yS_{\yq} \cond \dotso)\, \di\yq$' standing
  for appropriate singular measures.}

We can therefore analyse the context $\yI$ through the plausibility-\bd
indexed cir\-cum\-stances $\set{\yS^{(\yo)}_{\yq}}$, instead of the
$\set{\yC^{(\yo)}_j}$. The rationale behind this is that those
circumstances $\yC^{(\yo)}_j$ that belong to a given class $\yqc$ have all
the same effect on our judgement as regards the assignment and the update
of the plausibilities of the outcomes; moreover, the ratios of their
updated plausibilities $\pr(\yC^{(\yo)}_j \cond \dotso
\land\yI)$ cannot change upon acquisition of new measurement outcomes and
are equal to those of their prior plausibilities,
see~\citep{portamana2007b,portamana2007} and~\citep[remark~3 and
\sect~4.2]{portamanaetal2006}. It is therefore not unreasonable to handle
equivalent $\yC^{(\yo)}_j$ class-wise. The special notation chosen for the
different classes, `$\yS^{(\yo)}_{\yq}$', the index `$\yq$' in particular,
is to remind us on the grounds of what plausibility judgements (the $\yq$)
we grouped the original circumstances thus in the first place. But it is
only a notation, nothing more. Each $\yS^{(\yo)}_{\yq}$ is a disjunction of
propositions like, \eg, \prop{The coin is two-headed, or the tosses are
  made by a trickster with a predilection for heads, or \ldots}\ --- it is
\emph{not} a statement about the plausibilities $\yqq_i$ (nor about
`propensities').

The effect of this `bookkeeping' notation is, however, surprising for the
\emph{form} our induction
formulae~\eqref{eq:decomp_j}--\eqref{eq:upd_pr_Cj} take when expressed in
terms of the plausibility-indexed circumstances $\set{\yS^{(\yo)}_{\yq}}$.
The plausibility we assign to any collection of measurement outcomes takes
now, by \eqn~\eqref{eq:cond_plaus_Cq}, the form
\begin{multline}
  \label{eq:decomp_q}
  \pr( \underset{\text{\makebox[0pt]{$\yR_1$ appears
        $\yL_1$ times,
        \etc}}}{\underbrace{\yR^{(\yL)}_{i_{\yL}} \land
      \dotsb \land \yR^{(1)}_{i_{1}}}} \cond \yme \land  \yI) =
{}\\
\int
  \Bigl[\tprod_i \pr(\yR_i \cond \yM \land  \yS_{\yq} \land
  \yI)^{\yL_i} \Bigr] \, \pf(\yS_{\yq} \cond \yI) \,
  \di\yq \equiv
  \int \Bigl(\tprod_i \yqq_i^{\yL_i}\Bigr) \,
  \pf(\yS_{\yq} \cond \yI) \, \di\yq,
\end{multline}
with value numerically equal to that of
\eqref{eq:decomp_j}, and with the same remarks and
conventions about the index $\yo$ as made in
\sect~\ref{sec:intro_circ}. The expression above is
formally identical to the
de~Finettian's~\eqref{eq:deF_thm} with the correspondence
$\pf(\yS_{\yq} \cond \yI) \leftrightsquigarrow \yP(\yq
\cond \yI)$. In particular, the plausibility for the
outcome of any instance takes the form
\begin{equation}
  \label{eq:plaus_next_q}
  \pr(\yR^{(\yo)}_i \cond \yM^{(\yo)} \land  \yI) = 
\int \yqq_i \, \pf(\yS_{\yq} \cond \yI)  \, \di\yq,
\end{equation}
formally identical to~\eqref{eq:plaus_next}.

The plausibility conditional on the observation of $\yN$
outcomes takes the form
\begin{multline}
  \label{eq:new_ev_q}
  \pr(
\underset{\text{\makebox[0pt]{$\yR_i$ appears $\yL_i$ times}}}{\underbrace{
\yR^{(\yo_{\yN + \yL})}_{i_{\yN + \yL}} \land \dotsb \land
\yR^{(\yo_{\yN + 1})}_{i_{\yN + 1}} 
}}
\cond \yme \land
\underset{\text{\makebox[0pt]{$\yR_i$ appears $\yN_i$ times}}}{\underbrace{
\yR^{(\yo_{\yN})}_{i_{\yN}} \land \dotsb \land \yR^{(\yo_{1})}_{i_{1}} 
}}
\land \yI) 
={}
\\[2\jot]
\int \Bigl[\tprod_i  \pr(\yR_i \cond \yM \land  \yS_{\yq} \land
\yI)^{\yL_i} \Bigr] \, 
\pf(\yS_{\yq} \cond 
\yR^{(\yo_{\yN})}_{i_{\yN}} \land \dotsb \land \yR^{(\yo_{1})}_{i_{1}}
\land
\yI) \, \di\yq
\equiv{}
\\
\int \Bigl(\tprod_i \yqq_{i}^{\yL_i}\Bigr) \, \pf(\yS_{\yq}
\cond 
\yR^{(\yo_{\yN})}_{i_{\yN}} \land \dotsb \land \yR^{(\yo_{1})}_{i_{1}}
\land
\yI) \, \di\yq
\end{multline}
with
\begin{equation}
  \label{eq:upd_pr_Cq}
  \pf(\yS_{\yq} \cond   \yR^{(\yo_{\yN})}_{i_{\yN}} \land \dotsb \land \yR^{(\yo_{1})}_{i_{1}}
  \land\yI) \, \di\yq
  =
  \frac{
    \Bigl(\tprod_i \yqq_{i}^{\yN_i}\Bigr) \,
    \pf(\yS_{\yq} \cond   \yI) \, \di\yq
  }{
    \int \Bigl(\tprod_i \yqq_{i}^{\yN_i}\Bigr) \, 
    \pf(\yS_{\yq} \cond   \yI) \, \di\yq 
  },
\end{equation}
again formally identical to the
de~Finettian's~\eqref{eq:deF_new_ev} and~\eqref{eq:upd_Q}.

Apart from the congruence between their mathematical
\emph{forms}, the above formulae and those derived from
exchangeability have very different \emph{meanings}.
Whereas in the exchangeability approach the $\yq \equiv
(\yqq_i)$ were just parameters, in the Laplace-Jaynes
approach they are (numerical values of) \emph{actual
  plausibilities}, \viz\ the $\pr(\yR^{(\yo)}_i \cond
\yM^{(\yo)} \land \yC^{(\yo)}_j \land \yI)$ or the
$\pr(\yR^{(\yo)}_i \cond \yM^{(\yo)} \land \yS^{(\yo)}_{\yq}
\land \yI)$. Whereas in the exchangeability approach $\yq
\mapsto \yP(\yq \cond \dotso)$ was only a generalised
function, in the Laplace-Jaynes approach $\yq \mapsto
\pf(\yS_{\yq} \cond \dotso)$ is (the density of) an
\emph{actual plausibility distribution} --- a
distribution, however, not over `probabilities' or
`propensities', but rather over propositions like
\prop{The coin is two-headed, or the tosses are made by a
  trickster with a predilection for heads, or \ldots}.

Equations~\eqref{eq:new_ev_q} and~\eqref{eq:upd_pr_Cq} (as
well as~\eqref{eq:new_ev_j} and~\eqref{eq:upd_pr_Cj}) show
that the observation of measurement results has a double
`updating' effect in the circumstances approach. Not only
are the plausibilities of unobserved results updated, but
those of the circumstances as well; in fact, the former
updating happens through the latter. This is related to
what Caves calls `learning through a
parameter'~\citep{caves2000c}, although there are no
parameters here, only plausibilities. Also in the
exchangeability approach, one could argue, is the
generating function $\yP(\yq \cond \yI)$ updated; but its
updating only represents and reflects a change in our
uncertainty or degree of belief \emph{about the outcomes},
no more than that. In the Laplace-Jaynes approach, the
updating of $\pf(\yS_{\yq} \cond \yI)$ represents instead
a further change in our uncertainty about events or
phenomena other than the outcomes. We shall return to this
in a moment.

We finally also see how the Laplace-Jaynes approach
interprets and makes sense of the notions of `unknown
probability (or propensity)' and `probability of a
probability': what is unknown is not a probability, but
which circumstance from a set of empirical ones holds; the
second probability is therefore not about a probability,
but about an empirical circumstance.

\section{Limit for large number of observations}
\label{sec:large_N_q}

Let us suppose that the plausibility distribution
$\pf(\yS_{\yq} \cond \yI) \, \di\yq$ does not vanish for
any $\yq$. This implies that the set of `fine-grained'
circumstances $\set{\yC_j}$ is a continuum; with
analytical and topological care this case should not
present particular difficulties.

That assumption being made, from \eqn~\eqref{eq:upd_pr_Cq}
we have that as the number $\yN$ of observations increases,
the updated distribution for the plausibility-indexed
circumstances asymptotically becomes
\begin{equation}
  \label{eq:upd_PC_limit}
\pf(\yS_{\yq} \cond \yme \land
\underset{\text{\makebox[0pt]{$\yR_i$ appears $\yN_i$
      times}}}{\underbrace{
\yR^{(\yo_{\yN})}_{i_{\yN}} \land \dotsb \land \yR^{(\yo_{1})}_{i_{1}}
}}
  \land \yI)
\simeq
\delt\Bigl(\tfrac{\yNf}{\yN} - \yq\Bigr)
\quad \text{as $\yN \to \infty$,}
\end{equation}
which is formally identical to the
de~Finettian's~\eqref{eq:upd_Q_limit}. It also follows
that
\begin{equation}
  \label{eq:plaus_next_large_N_circ}
  \pr(\yR^{(\yo_{\yN+1})}_i \cond  \yM^{(\yo_{\yN+1})}  \land
\underset{\text{\makebox[0pt]{$\yR_i$ appears $\yN_i$
      times}}}{\underbrace{
\yR^{(\yo_{\yN})}_{i_{\yN}} \land \dotsb \land \yR^{(\yo_{1})}_{i_{1}}
}}
  \land\yI) 
\simeq \frac{\yN_i}{\yN}
\quad \text{as $\yN \to \infty$,}
\end{equation}
exactly as in~\eqref{eq:plaus_next_large_N}. Also in the Laplace-Jaynes
approach, then, the plausibilities assigned to unobserved outcomes get
nearer their observed relative frequencies as the number of observations
gets larger (under the assumptions specified above). And persons sharing
the same data and having compatible initial plausibility assignments tend
to converge to similar plausibility assignments, as regards their
respective sets of plausibility-indexed circumstances (and as regards
unobserved measurement outcomes). \Cf\ the discussion in
\sects~\ref{sec:sureness} and~\ref{sec:when_which}, and see also
Jaynes~\citep[\chap~18]{jaynes1994_r2003} (see
also~\citep[lect.~18]{jaynes1954}\citep[lect.~5]{jaynes1959}).

What happens if the distribution $\pf(\yS_{\yq} \cond \yI)
\, \di\yq$ vanishes at, or in a neighbourhood of, the
point $(\yqq_i)=(\yN_i / \yN)$, as may be the case when
the number of `fine-grained' circumstances $\set{\yC_j}$ is
finite? It happens that the updated distribution gets
concentrated at those $\yS_{\yq}$ (and those $\yC_j$) with
$\yq$ (respectively $(\yqq_{ki})$) nearest $(\yN_i / \yN)$
and for which the initial plausibility does not vanish.
One should make the adjective `nearest'
topologically more precise. There are also interesting
results at variance with the asymptotic
expression~\eqref{eq:plaus_next_large_N} when $(\yN_i /
\yN)$ cannot be obtained as a convex combination of those
$\yq$ (respectively $(\yqq_{ki})$) whose related
plausibilities do not vanish. All this is left to another
study.

\chapter{Discussion}
\label{sec:discu}

\section{Relations between the circumstance and
  exchangeability approaches}
\label{sec:relat_c_e}

The Laplace-Jaynes approach and the infinite-\bd
exchangeability one do not exclude each other and may be
used simultaneously in many problems. In fact, if in a
given problem we can introduce `circumstances', and the
number of measurement instances is potentially infinite,
the resulting distributions for collections of outcomes
are then infinitely exchangeable and all the results and
representations based on exchangeability also apply,
beside those based on the circumstance representation.
This leads to the following powerful proposition:
\begin{corol}
In the Laplace-Jaynes approach, if the number of
measurement instances is potentially infinite then
\begin{multline}
  \label{eq:corol1}
  \pr(
\underset{\text{\makebox[0pt]{$\yR_1$ appears $\yL_1$ times, \etc}}}{\underbrace{\yR^{(\yo_{\yL})}_{i_{\yL}} \land \dotsb \land  \yR^{(\yo_{1})}_{i_{1}}}}
\cond \yme \land  \yD \land \yI) =
{}\\
  \int \Bigl(\tprod_i \yqq_i^{\yL_i}\Bigr) \, \yP(\yq \cond \yD \land \yI) \,
  \di\yq
=
\int \Bigl(\tprod_i \yqq_i^{\yL_i}\Bigr) \,
  \pf(\yS_{\yq} \cond \yD \land \yI) \, \di\yq,
\end{multline}
and in particular
\begin{equation}
  \label{eq:corol2}
  \yP(\yq \cond \yD \land \yI) \,  \di\yq
=
\pf(\yS_{\yq} \cond \yD \land \yI) \, \di\yq,
\end{equation}
for any $\yL$, $\set{i_\yj}$, and any $\yD$ representing a
(possibly empty) conjunction of 
measurement outcomes not containing the instances $\yo_1,
\dotsc, \yo_\yL$.
\end{corol}
The equalities of the above Proposition may be read in two
senses, whose meaning is the following: If in a given
situation we have introduced circumstances
$\set{\yS_{\yq}}$ (or, equivalently, $\set{\yC_{j}}$) and
assigned or calculated their plausibility density
$\pf(\yS_{\yq} \cond \yD \land \yI)$, then we also know,
automatically, the generating function $\yP(\yq \cond \yD
\land \yI)$ of de~Finetti's representation theorem, which
we should have introduced had we taken an exchangeability
approach.
Vice versa, if we approach the problem through exchangeability and assign
an infinitely exchangeable distribution to the possible infinite
collections of outcomes, we have automatically assigned also the
plausibility density $\pf(\yS_{\yq} \cond \yD \land \yI)$ for
\emph{whatever} set of plausibility-indexed circumstances $\set{\yS_{\yq}}$
we might be willing to introduce. The density of the Laplace-Jaynes
approach and the generating function of the exchangeability one are, in
both cases, numerically equal.

What is remarkable in the second direction of the Proposition is that the
plausibilities of the plausibility-indexed circumstances are completely
determined from our exchangeable plausibility assignment, even when we have
not yet specified what those propositions are about! Although remarkable,
this fact is a consequence of the particular way the plausibility-indexed
circumstances are formed from the `fine-grained' ones. Note, moreover, that
the plausibilities of the `fine-grained' circumstances $\set{\yC_{j}}$ would
not be completely determined in general, even though their values would be
constrained by \eqn~\eqref{eq:plaus_Cq_from_Cj}.

Some philosophical importance has the fact that
$\pf(\yS_{\yq} \cond \dotso)$, numerically equal to
$\yP(\yq \cond \dotso)$, is an `actual' plausibility
distribution, whereas the latter is only a generating
function. It is practically impossible to specify most
infinitely exchangeable plausibility distributions
\emph{directly},
and this is one of the main points and advantages of
de~Finetti's theorem: a de~Finettian can specify an
infinitely exchangeable distribution by making a (often
necessary) detour and specifying $\yP$ instead. It is,
however, a bit disconcerting the fact that one is almost
always forced to specify the `as if' instead of what
according to de~Finetti is `the actual plausibility'. The
Laplace-Jaynes approach gives instead a direct meaning to
the generating function $\yP$ \emph{as a plausibility
  distribution}, so that one needs not feel embarrassed to
specify it.

Can all situations which can be approached through
infinite exchangeability also be approached from the
Laplace-Jaynes point of view? Asking this means asking
whether all situations for which we deem exchangeability
to apply can be analysed into circumstances having the
properties~\eqref{eq:exc_exh}--\eqref{eq:induction}. There
is surely much place for discussion on the answer to this
question, if by `circumstances' we really mean, in some
sense, `interesting circumstances'. One could also wonder
whether on a formal level the answer could be `yes'. The
reason is that the Laplace-Jaynes approach would seem to
formally include the infinite-exchangeability one. To see
how, make first a judgement of infinite exchangeability,
and then introduce and define a set $\set{\yC_{\yf}}$ of
propositions stating that the limiting relative
frequencies of an infinite collection of outcomes have
certain values $\yf$. This should be possible since the
existence of this limit under exchangeability is
guaranteed by de~Finetti's theorem. 
Intuitively, such a set $\set{\yC_{\yf}} \defs
\set{\yC^{(\yo)}_{\yf}}$ for any $\yo$ would seem to
constitute (under a judgement of infinite exchangeability)
a set of circumstances that satisfy
properties~\eqref{eq:exc_exh}--\eqref{eq:induction}. If
this were true then, whenever infinite exchangeability
applied, the Laplace-Jaynes approach would be at least
formally viable. But intuition is not a reliable guide
here (the definition of a proposition like $\yC_{\yf}$,
\eg, would apparently involve a disjunction over the set
of permutations of natural numbers), and the above
reasoning is not mathematically rigorous.
We have not the mathematical knowledge necessary to
rigorously analyse and answer this sort of conjecture, and
since we find it in any case uninteresting, let us not
discuss it any further. It can be added, however, that any
criticism from de~Finettians with regards to the
infinities involved in the argument would be like sawing
the branch upon which they are sitting themselves, since
we do not see how one can effectively make a judgement of
infinite exchangeability without first considering
possibly problematic propositions like $\set{\yC_{\yf}}$.

\section{`Unsure' and `unstable' plausibiliy assignments}
\label{sec:sureness}

The presence of circumstances in the analysis of the
problem, and the consequent fact that the plausibility of
the outcomes can be decomposed as in
\eqn~\eqref{eq:plaus_next_q}, suggest an interpretation of
those feelings of `uncertainty' and `stability' about a
plausibility assignment mentioned in
\sect~\ref{sec:why_appr}, point~\eqref{item:uncertain}. A
person might feel `sure' about an assignment $1/2$ to
heads in the situation $\yIs$ because of the following,
perhaps unconscious, reasoning: `If I knew that the coin
was two-headed or the toss method favoured heads, I'd
assign $1$ to heads. If on the contrary I knew that the
coin was two-tailed or the toss method favoured tails, I'd
assign $0$ to heads. If I knew that the coin was a usual
one and the tosser knew nothing about tossing methods, I'd
assign $1/2$ to heads. But the coin, although I gave to it
a rapid glance only, seems to me quite common and
symmetric; and I know that the tosser, a friend of mine,
knows no tossing tricks. So I can safely exclude the first
two possibilities, and I'm practically sure of the third.
Yes, I'll give $1/2$ to heads'. In this reasoning, the
person entertains (for each $\yo$) a set of possibilities,
like \prop{The coin is two-headed}, \prop{The toss method
  favours tails}, \etc{} From these a set of three
plausibility-indexed circumstances $\yS_{\yq}$, with $\yq
\equiv (\yqq_\text{heads},
\yqq_\text{tails})$,\footnote{Note, once more, that the
  $\yS_{\yq}$ concern not plausibilities but statements
  like, in this case, \prop{The coin is a usual one and
    the tosser knows nothing about tossing methods},
  \etc}\ is distilled, corresponding to the $\yq$ values
$(1,0)$, $(1/2, 1/2)$, and $(0,1)$. To these circumstances
the person assigns, given the background knowledge $\yIs$,
the plausibilities
\begin{equation}\label{eq:Pq_sure}
  \begin{split}
    \pr(\yS_{(1,0)} \cond \yIs) &\approx 0, \\
    \pr(\yS_{(1/2,1/2)} \cond \yIs)&\approx 1, \\
    \pr(\yS_{(0,1)} \cond \yIs) &\approx 0.
  \end{split}
\end{equation}
Then, according to the rules of plausibility theory, the
total plausibility assigned to heads is
\begin{equation}
  \label{eq:example_sure}
  \begin{split}
    \pr(\prop{heads} \cond \yIs) &= \tsum_{\yq}
    \pr(\prop{heads} \cond \yS_{\yq} \land \yIs) \,
    \pr(\yS_{\yq} \cond \yIs),
    \\
    &=1 \times \pr(\yS_{(1,0)} \cond \yIs) + 1/2 \times
    \pr(\yS_{(1/2,1/2)} \cond \yIs) + 0 \times
    \pr(\yS_{(0,1)} \cond \yIs) ,
    \\
    &\approx 1 \times 0 + 1/2 \times 1 + 0 \times 0 = 1/2.
  \end{split}
\end{equation}
The fact to notice here is that amongst all the
circumstances, those that lead to a conditional
plausibility for heads near to the total plausibility,
\viz~$1/2$, have together far higher plausibility than the
others. This fact can be interpreted as the source of the
feelings of sureness and stability associated to that
final plausibility assignment: the person has conceived a
number of hypotheses, and the total plausibility assigned
on the grounds of these is practically equal to the
plausibilities assigned conditionally on the most
plausible hypotheses only. Put it otherwise, the most
plausible circumstances are not discordant, all point more
or less to the same plausibility assignment.
Mathematically this is reflected in the fact that the
distribution $\pr(\yS_{\yq} \cond \yIs)$ is concentrated
around the median.

The reasoning of a person who feels `unsure' about an
assignment $1/2$ to heads in the situation $\yIu$ might go
instead as follows: `Had I known that the coin was a
common one, I'd have assigned $1/2$ to heads. But I have
heard from reliable sources that this coin is not a common
one, so I can safely exclude that possibility. If I knew
that the coin was two-headed, I'd assign $1$ to heads. If
I knew that the it was two-tailed, I'd assign $0$ to
heads. But I'm completely unsure about these two
possibilities! I have to give $1/2$ to heads then'. Also
in this reasoning the person introduces a collection of
hypotheses and from these distills a set of three
plausibility-indexed circumstances similar to those
introduced by the `sure' person. It is in the assessment
of the plausibilities $\pr(\yS_{\yq} \cond \yI)$ that the
two persons --- owing to their different background
knowledges $\yIs$, $\yIu$ --- differ. In this case the
assessment yields
\begin{equation}\label{eq:Pq_unsure}
  \begin{split}
    \pr(\yS_{(1,0)} \cond \yIu) &\approx 1/2, \\
    \pr(\yS_{(1/2,1/2)} \cond \yIu)&\approx 0, \\
    \pr(\yS_{(0,1)} \cond \yIu) &\approx 1/2,
  \end{split}
\end{equation}
and therefore the unsure person assign to the outcome
`heads' the total plausibility
\begin{equation}
  \label{eq:example_unsure}
  \begin{split}
    \pr(\prop{heads} \cond \yIu) &= \tsum_{\yq}
    \pr(\prop{heads} \cond \yS_{\yq} \land \yIu) \,
    \pr(\yS_{\yq} \cond \yIu),
    \\
    &=1 \times \pr(\yS_{(1,0)} \cond \yIu) + 1/2 \times
    \pr(\yS_{(1/2,1/2)} \cond \yIu) + 0 \times
    \pr(\yS_{(0,1)} \cond \yIu),
    \\
    &\approx 1 \times 1/2 + 1/2 \times 0 + 0 \times 1/2 =
    1/2.
  \end{split}
\end{equation}
The final assignment is the same as that of the sure
person, but the most plausible circumstances for the
unsure person would yield conditional plausibilities not
near to the final value~$1/2$. The unsure person would
have given $0$ or $1$ if a little bit more knowledge of
the situation had been available.\footnote{The assignment
  will in fact collapse onto one of those two values as
  soon as one toss is observed, leading to nought-or-one
  updated plausibilities for the circumstances
  $\yS_{(0,1)}$ and $\yS_{(1,0)}$.} In other words, the
most plausible circumstances are discordant and point to
different plausibility assignments, and this can be
interpreted as the source of the feelings of unsureness
and instability. In mathematical terms, the distribution
$\pr(\yS_{\yq} \cond \yIu)$ is not concentrated around the
median.

The two examples above are very simple, the number of possible
plausibility-indexed circumstances being only three. With a deeper analysis
this number could be very large, and the assignments~\eqref{eq:Pq_sure}
and~\eqref{eq:Pq_unsure} would be replaced by effective continuous
distributions like \eg
\begin{align}
\pf(\yS_{\yq} \cond \yIs) \,\di\yq &\propto 
(\yqq_\text{heads})^{20} \, (\yqq_\text{tails})^{20} \,
\delt(1-\yqq_\text{heads}-\yqq_\text{tails})\,\di\yq,
\label{eq:quant_sure}
\\
\pf(\yS_{\yq} \cond \yIu)\,\di\yq &\propto 
(\yqq_\text{heads})^{-39/40} \, (\yqq_\text{tails})^{-39/40} \,
\delt(1-\yqq_\text{heads}-\yqq_\text{tails})\,\di\yq,
\label{eq:quant_unsure}
\end{align}
whose graphs are those of Fig.~\ref{fig:coins}, blue
dashed curve for the first and orange continuous curve for
the second. These are Dirichlet (or beta) distributions,
known for their particular properties, for which see
\eg~\citep{johnson1932c,good1965,novick1969,jaynes1986d_r1996,bernardoetal1994}
(\cf\ also~\citep{lindleyetal1976}).

The degree of `sureness' or `stability' about a plausibility assignment can
thus be apparently captured, formalised, and even quantified by the
plausibility distribution of a set of circumstances, $\pr(\yS_{\yq} \cond
\yI)$, and by its `width' in particular, which can be defined and
quantified in a different number of ways (entropy, standard deviation,
\etc). Graphs such as those of Fig.~\ref{fig:coins} have then a legitimate
and quantitative meaning within (single-valued) plausibility theory: they
do not represent `probabilities of probabilities'; they represent
plausibility distributions amongst different unknown empirical
circumstances, grouped for simplicity according to the conditional
plausibilities they lead to.

This interpretation is also consonant with the fact that,
according to the results of \sect~\ref{sec:large_N_q},
\eqn~\eqref{eq:upd_PC_limit}, as the observed data
accumulate the distribution $\pr(\yS_{\yq} \cond \yI)$
gets more and more peaked around a particular value which
gets also nearer the median, reflecting the fact that
accumulation of data tends to make us surer of our
plausibility assignments.

For further discussion and references, see Jaynes
\citetext{\citealp[\chap~18]{jaynes1994_r2003}; \cf\
  also~\citealp[lect.~18]{jaynes1954}; \citealp[lect.~5]{jaynes1959}}
and~\citep{portamanaetal2006}.

\section{Finite and small number of possible observations}
\label{sec:finite_obs}

As mentioned in \sect~\ref{sec:why_appr} with regard to
point~\eqref{item:finite_case}, there are situations in
which the infinite-\bd exchangeability approach cannot be
adopted: situations where we know that the number of
possible measurement instances is finite --- and, in
particular, small. In this case a de~Finettian has three
possible options as regards the application of
exchangeability.

The first is to make a judgement of \emph{finite}
exchangeability and use the related representation
theorem. It is known that the representation theorem
for finite exchangeability has a different form from that
for infinite exchangeability, \eqn~\eqref{eq:deF_thm}.
The plausibility for a collection of outcomes takes in the
finite case the form of a mixture of hypergeometric
distributions (`urn samplings without replacement'). In
the case of a finite but large number of possible
observations this representation converges to the
infinitely exchangeable
one~\citep{diaconis1977,diaconisetal1980}\citep[see
also][]{zabell1989}, so the de~Finettian can make sense of
the propensitor's `unknown propensity' reasoning and of
the related formulae, which are formally identical
with~\eqref{eq:deF_thm}--\eqref{eq:upd_Q}, as
approximations. But in the case of a small maximum number
of possible observations the discrepancy between the two
representations is too large, and if the `propensitor'
obstinately uses the `unknown propensity' formulae, the
de~Finettian will then not be able to `make sense' of
them, as instead was the case with infinite
exchangeability. 

The second option is to use de~Finetti's theorem as
generalised by Jaynes~\citep{jaynes1986c}, \ie\ extended
to generating functions of any sign, but restricting the
choice of the latter to positive ones only. But this
restriction would lack meaningful motivation: What kind of
plausibility judgement would the restriction reflect? ---
For we must remember that the specification of the
generating function is for the de~Finettian only a detour
to specify the plausibility distribution of the outcomes,
which is the only `actual' one. The restriction to
positive generating functions would only seem to reflect
the third option, to which we now turn.

The third option is to enlarge the finite collection of
measurement instances with an infinite number of fictive
ones, and then make a judgement of infinite
exchangeability for the enlarged collection. With this
artifice the formulae of the propensitor can be recovered,
and seemingly with a meaningful motivation. It seems to
us, however, that the introduction of very many ($\infty$)
fictive measurement instances, apart from being
unpleasant, is formally inconsistent. For the fact that
one \emph{knows} that the number of measurement instances
can but be finite, say at most $\yn$, must be included in
the context $\yI$; \viz, $\yM^{(\yo)} \land \yI$ is false
for $\yo > \yn$. The plausibility of the outcomes of the
fictive `additional' ($\yo > \yn$) measurements, which is
conditional on $\yI$, is then by definition nought, or
undefined. One cannot even say `Let us suppose that
additional measurements were possible', because such a
proposition conjoined with $\yI$ would yield a false
context and the plausibilities conditional on it would
then be undefined. An alternative would be to eliminate
from the context $\yI$ those `elements' that bound the
number of measurement instances, so that the
plausibilities of the fictive additional measurements
would neither be nought nor undefined in the new,
mutilated context. But mutilation of prior knowledge ---
granted its feasibility --- can be dangerous.

The Laplace-Jaynes approach, on the other hand, applies
unaltered to the case of a finite, even small, maximum
number of measurement instances. This is clear if we look
again at
properties~\eqref{eq:screening}--\eqref{eq:induction}:
they nowhere require the instance index $\yo$ to run to
infinity, as cursively remarked directly after their
enunciation.
Hence, through the Laplace-Jaynes approach we can make
straightforward sense of the propensitor's procedure also
in these finite situations.

\section{When is the Laplace-Jaynes approach useful? Making
  allowance for the grounds behind exchangeability
  assignments: some examples}
\label{sec:when_which}


We have seen in \sect~\ref{sec:relat_c_e} that the two
approaches are not mutually exclusive but can coexist and
strongly `interact'. But in which sense are the two
approaches complementary? How ought we to decide which
approach to choose? Why? There is no definite answer: as
de~Finetti warns~\citep[\sect~20]{definetti1931_t1989}:
\begin{quote}
  in certain enterprises it can be better to evaluate the
  probability of favourable outcomes all in a block, to
  see at a glance whether the investment is secure or
  insecure, and in others it is better to reach this
  conclusion starting from an analysis of the individual
  factors that are in play. There is no conceptual
  difference between the two cases. Someone who wants to
  estimate the area of a rectangular field can with the
  same right estimate the area directly in hectares, or
  estimate the lengths of the sides and multiply them:
  reasons of convenience, practicality and custom will
  make one method preferable to the other, the one that
  seems to us more trustworthy in relation to our capacity
  to judge, or we can follow both methods, or try other
  ones, and consider all of them in fixing our opinion.
\end{quote}
In assigning a plausibility to some measurement outcomes
we may simply want, depending on `convenience,
practicality and custom', to assign plausibilities to all
possible collections of similar outcomes (indirectly, by
specifying the generating function of de~Finetti's
theorem), update the plausibilities according to
observations already made, and then marginalise for the
instances of interest, as described in
\sect~\ref{cha:appr_inf_exch}. Or we may want to analyse
the contexts of the measurements into more details and
assign plausibilities to these, obtaining the final
plausibilities for the outcomes of interest by the theorem
on total probability, as described in
\sect~\ref{sec:exch_wh}. Neither approach is more
`objective' than the other, for both must start from some
set of (subjective, the de~Finettian says) plausibilities
that must be given without further analysis.

In situations such as that of a `common' coin toss or die
throw, we (the authors) usually prefer in general the
approach through exchangeability. Although such situations
could be analysed into more details and hypotheses about
the way of tossing \etc, we are not really interested in
these and prefer to make an initial exchangeable
plausibility assignment for future tosses, that we shall
update according to observations made --- the number of
which can be very large (\ie, infinite exchangeability can
be applied, at least as an approximation). One could say
that it is the plausibilities `$\pr(\yR^{(\yo)}_i \cond
\dotso)$' that are of interest, and we collect measurement
data to be put in place of the dots in order to update
those plausibilities. But there are examples, especially
in the study of natural phenomena, where it is the details
--- the circumstances --- that interest us most and that
we want to investigate. In these cases measurement
outcomes are collected not only for the sake of prediction
of other, unknown ones, but also in order to modify our
uncertainty about the circumstances, \ie, to change our
initial plausibility assignments about them. In other
words, the plausibilities `$\pr(\yC_j \cond \dotso)$'
interest us as much as the `$\pr(\yR^{(\yo)}_i \cond
\dotso)$', and measurement data are collected to update
both.

A powerful feature of the approach through plausibility-indexed
circumstances is that you can `wait' to think about and to explicitly
define the circumstances of interest until you have collected a large
amount of observations. The `procedure' is as follows:

\begin{enumerate}
  \item\label{item:ass_num_C}\emph{Imagine} to have introduced numerous
  circumstances $\set{\yC_j}$, to have assigned the plausibilities
  $\pr(\yR_i \cond \yM \land \yC_j \land \yI)$, and to have performed the
  `coarse-graining' of the circumstances into the set $\set{\yS_{\yq}}$.

  \item\label{item:ass_smooth_p}Assume to have assigned plausibilities
  $\pr(\yC_j \cond \yI)$ such that the plausibility density $\pf(\yS_{\yq}
  \cond \yI)$ over the plausibility-indexed circumstances is enough smooth
  and non-\bd vanishing at every point. How much is `enough' is related to
  the size of the collected data as for the next step.

\item Collect a large amount of outcome observations $\yD$. What counts as
  `large' is related to the smoothness assumed in the previous step for the
  plausibility density over the plausibility-indexed circumstances.

  \item\label{item:update_singles} Update your distributions according to
  the observations $\yD$. If $\yf \equiv (\yff_i)$ are the relative
  frequencies of the observed outcomes, the updated plausibility
  distribution over the plausibility-indexed circumstances would now be
  concentrated around the circumstance labelled by $\yf$, \ie,
  $\pf(\yS_{\yq} \cond \yD \land \yI)\, \di\yq \approx \delt(\yq - \yf)\,
  \di\yq$; \cf\ \sect~\ref{sec:large_N_q}. This means that you would judge
  $\yS_{\yf}$ to be the most plausible circumstance given the evidence
  $\yD$. (But remember: the introduction of circumstances is up to now only
  imagined.)

\item Now re-examine the context as it was before the
  observations, and \emph{really} introduce a set of
  circumstances of interest, respecting the assumptions in
  steps~\ref{item:ass_num_C} and~\ref{item:ass_smooth_p}.
  In this process, however, you only need to concentrate
  on those circumstances conditional on which the outcomes
  of any measurement instance are near to $\yf$, \ie\
  those for which $\pr(\yR_i \cond \yM \land \yC_j \land
  \yI) \approx \yff_i$. This restriction is sensible
  since, as for step~\ref{item:update_singles} above,
  these are the circumstances that have now become most
  plausible in the light of the observed data.
\end{enumerate}
This process reflects \eg\ what we do when, having tossed
a coin many times and seen that we nearly always obtain
heads, we say `there's something peculiar going on here,
let's see what it can be', and begin to look for
circumstances (like who tossed the coin, how the coin was
tossed, on which surface it was tossed, how it was
manufactured, \etc)\ which presumably lead to the observed
peculiar behaviour.


Examples in the same spirit, although more complex, abound
in the natural sciences. Consider the following example
from the study of granular
materials~\citep[\eg,][]{jaegeretal1996,degennes1999}.
These are made of a very large number of macroscopic
(thermal agitation is unimportant) but small particles,
that are hence considered point-like. These materials can
then be studied by the methods of statistical mechanics.
Often of interest is the particle motion resulting from an
externally applied force and the particles' collisions. In
the study by Rouyer and Menon~\citep{rouyeretal2000} (to
which we refer for details) we are interested in the
measurement --- our $\yM$ --- of the horizontal velocity
component $\yv$ of any particle. The possible measurement
outcomes $\set{\yR_\yv}$ are the possible different values
of $\yv$.\footnote{Hence, the set of outcomes is
  continuous here but, with appropriate mathematical care,
  this is not so important.} What is the plausibility
distribution that we assign to the different outcomes?
Depending on the circumstances, we should make different
assignments. For example, if we knew that the particle
collisions were elastic, the particle density enough low,
long-range interactions negligible, and a couple more of
details --- all of which we represent by $\yCmb$ ---, we
should assign a Maxwell-Boltzmann, Gaussian distribution
$\pf(\yR_\yv \cond \yM \land \yCmb \land \yI) \, \di\yv
\propto \exp(\yv^2/ \yvo^2) \, \di\yv$, with an
appropriate constant $\yvo$. (In the present case we know
the collisions to be inelastic, so that $\yCmb$ is from
the beginning known to be false; but let us disregard this
to make the example more interesting.) If we
knew~\citep{rouyeretal2000} that the collisions were
inelastic, knew that the density presented
inhomogeneities, judged the particle density very relevant
for the velocity distribution, and knew some other details
--- denote all them by $\yCpu$ ---, we should then assign
a particular non-Gaussian plausibility distribution,
according to the kinetic-theoretical considerations of
Puglisi \etal~\citep{puglisietal1999}. Finally, if we
knew~\citep{rouyeretal2000} that the collision were
inelastic, that energy was injected in the system
homogeneously in space, and some other details --- call
them $\yCne$ ---, we should then assign another particular
non-Gaussian plausibility distribution, according to the
study by van Noije and Ernst~\citep{vannoijeetal1998}. (In
the present case we know that energy is injected from the
boundary, so that $\yCne$ is known to be false; again, we
disregard this.)

We do not consider other possible circumstances for the
moment, but go on to collect a large amount of
observations instead --- call these $\yD$. The
distribution of frequencies thus observed goes like
$\yN_\yv/\yN \, \di\yv = \exp(\yv^\yal/ \yvu^\yal)\,
\di\yv$ with $\yal \approx 3/2$, a distribution very
similar to that assigned on the grounds of $\yCpu$, and
partially similar to that assigned conditional on $\yCne$.
By \eqn~\eqref{eq:plaus_next_large_N} we are thus led, in
both the exchangeability and the Laplace-Jaynes approach, to
assign
\begin{gather}
  \label{eq:grain_large_N}
  \pf(\yR^{(\yN+1)}_\yv \cond \yM^{(\yN+1)} \land \yD \land
  \yI)\, \di\yv
\approx
\exp(\yv^\yal/ \yvu^\yal)\, \di\yv \\
\intertext{and to say that}
\label{eq:gen_fun_grain_N}
\yP(\prop{$\exp(\yv^\yal/ \yvu^\yal)$} \cond \yD \land
\yI) =
\pf(\yS_{\prop{$\exp(\yv^\yal/ \yvu^\yal)$}} \cond \yD \land
\yI) \quad\text{has a very large value}.
\end{gather}
You note that $\yP$ has \prop{$\exp(\yv^\yal/ \yvu^\yal)$}
as argument, which is also the index of $\yS$. This is
because $\yv$ is a continuous variable, hence the vectors
$\yq \equiv (\yqq_i)$ of our previous discussions become
\emph{functions} $\yg(\yv)$ here. The function
\prop{$\exp(\yv^\yal/ \yvu^\yal)$} thus represents a
particular \emph{value} of what is denoted by `$\yq$' in
\sects~\ref{cha:appr_inf_exch} and~\ref{sec:exch_wh}.

The analysis has up to now involved both the
exchangeability and the Laplace-Jaynes approaches, and at
this point the exchangeability approach basically
terminates. But the interesting part of the Laplace-Jaynes
approach, instead, begins. In fact, from
\eqn~\eqref{eq:gen_fun_grain_N} we are led to assign a
vanishing plausibility to the circumstance $\yCmb$ and a
small one to $\yCne$ (which we already knew to be false,
however) and a non-negligible plausibility to the
circumstance $\yCpu$. This prompts other studies and
experiments in order to update that plausibility so as to
possibly decrease our uncertainty. Rouyer and
Menon~\citep{rouyeretal2000} actually do this, and
eventually come to a vanishing plausibility for the
circumstance $\yCpu$. Other circumstances are still to be
formulated and introduced.

In the above typically physical example the
`circumstances' are indeed intended as `causes' or
`mechanisms'. But as we remarked in
\sect~\ref{sec:intro_circ}, this needs not always be the
case. A very interesting and curious example, for which we
believe the circumstances cannot be interpreted as
`causes' or `mechanisms', is provided by the study of the
Newcomb-Benford
law~\citep{newcomb1881,benford1938,raimi1976,raimi1985},
here tersely described by Raimi~\citep[p.~521]{raimi1976}:
\begin{quotation}
  It has been known for a long time that if an extensive
  collection of numerical data expressed in decimal form
  is classified according to first significant digit,
  without regard to position of decimal point, the nine
  resulting classes are not usually of equal size. Indeed,
  [\ldots] for the occurrence of a given first digit $i$ 
  ($i=1, 2, \dotsc, 9$),\footnote{`$p$' in Raimi's text.}
  many observed tables give a frequency approximately
  equal to $\log_{10}[(i+1)/i]$. Thus the initial digit
  $1$ appears about $.301$ of the time, $2$ somewhat less
  and so on, with $9$ occurring as a first digit less than
  $5$ percent of the time. (We do not admit $0$ as a
  possible first digit.)\par
  This particular logarithmic distribution of first
  digits, while not universal, is so common and yet so
  surprising at first glance that it has given rise to a
  varied literature, among the authors of which are
  mathematicians, statisticians, economists, engineers,
  physicists and amateurs.
\end{quotation}
This example can be analysed along the lines of the
previous one. The `measurement' $\yM$ is the observation
of the first digits in a given collection of numerical
data, the possible `outcomes' being $\yR_1, \dotsc,
\yR_9$. Having made a judgement of exchangeability as
regards the outcomes of any number of
observations,\footnote{Since the number of data is finite,
  infinite exchangeability can here be used only as an
  approximation or an idealisation.} after many
observations our plausibility assignment will be very near
to the observed frequencies, in this case approximately
$\bigl(\log_{10}[(i+1)/i]\bigr)$. Here the approach
through exchangeability stops. But the Laplace-Jaynes
approach does not. According to
\eqn~\eqref{eq:upd_PC_limit} the circumstances that have
acquired highest plausibility are those which lead us to
assign the `surprising' distribution
$\bigl(\log_{10}[(i+1)/i]\bigr)$. The `varied literature'
which Raimi mentions consists almost exclusively in
searches and studies of such circumstances. The
significant point is that many proposed ones concern not
`causes' or `mechanisms', but symmetries.

A general and more thorough discussion of how the
Laplace-Jaynes approach fits into the formalism of
statistical physics will be given in our next study (in
preparation), together with an analysis of
point~\eqref{item:quantum} of \sect~\ref{sec:why_appr}.

\chapter{Generalisations}
\label{cha:general}

In the Laplace-Jaynes approach introduced in
\sect~\ref{sec:intro_circ} a set of possible circumstances
is defined for each measurement instance, and there is a
mutual correspondence of these sets, mathematically
expressed in particular by \eqns~\eqref{eq:ind_inst}
and~\eqref{eq:induction}. These express the idea that
there is a similarity amongst the measurement instances,
for they can be analysed into the `same' set of
circumstances; and also the idea that the `same', but
unknown, circumstance holds in all instances. This
approach can be generalised in different directions.

A first generalisation is to introduce a set of
circumstances $\set{\yC_{\yl}}$, not for each measurement
instance, but for all of them \emph{en bloc}. \Ie, each
circumstance $\yC_{\yl}$ concerns \emph{all} measurement
instances. This idea can be easily illustrated if
different measurement instances are characterised by
different times: then each $\yC_{\yl}$ can represent a
possible `history' of the details of the
instances.\footnote{\label{fn:explan_hist}Note, however,
  that different measurement instances do not need to be
  associated to different times in general, so the
  `history interpretation' is just an example. A more
  general way to think of the circumstances
  $\set{\yC_{\yl}}$ is to introduce for each measurement
  instance $\yo$ a set of circumstances $\set{\yC^{(\yo)}_j}$,
  as in \sect~\ref{sec:intro_circ}, but without assuming
  any correspondence amongst the sets, not even the same
  cardinality. We then consider all possible conjunctions
  $\Land_{\yo} \yC^{(\yo)}_{j_{\yo}}$, and each such conjunction can
  be taken to be by definition one of the $\yC_{\yl}$.}
Modifying
properties~\eqref{eq:exc_exh}--\eqref{eq:ind_inst} and the
subsequent sections by the formal substitution
$\yC^{(\mathord{\cdot})}_{j} \rightsquigarrow \yC_{\yl}$,
and dropping property~\eqref{eq:induction},
most part of the analysis and discussion presented in the
previous sections holds unchanged, or with minor
adaptations, for this generalisation.

A second generalisation, with similarities with the
preceding one, is made by dropping
property~\eqref{eq:induction} only. This means that we do not
assume the `same' circumstance to hold at every
measurement instance. All conjunctions $\Land_\yj
\yC^{(\yj)}_{j_\yj}$, where the $j_\yj$ can differ for different
$\yj$, can then have non-vanishing plausibilities (these
conjunctions are obviously similar to the circumstances
$\yC_{\yl}$ above; \cf\ footnote~\ref{fn:explan_hist}).
With this generalisation formulae~\eqref{eq:decomp_j}--%
\eqref{eq:upd_pr_Cj} do not hold; in their place we have
\begin{multline}
  \label{eq:decomp_j_new}\tag{\ref{eq:decomp_j}$'$}
  \pr(
\yR^{(\yo_{\yL})}_{i_{\yL}} \land \dotsb \land
\yR^{(\yo_{1})}_{i_{1}}
\cond \yme \land  \yI) =
{}\\
\sum_{j_1, j_2, \dotsc} 
 \pr(\yR^{(\yo_1)}_{i_1} \cond \yM^{(\yo_1)} \land  
\yC^{(\yo_1)}_{j_{\yo_1}} \land \yI)
\dotsm
 \pr(\yR^{(\yo_\yL)}_{i_\yL} \cond \yM^{(\yo_\yL)} \land  
\yC^{(\yo_\yL)}_{j_{\yo_\yL}} \land \yI)
\, \pr\bigl(\tland_\yj \yC^{(\yj)}_{j_\yj} \cond \yI\bigr)
\equiv
{}\\
\sum_{j_1, j_2, \dotsc}
 \yqq_{i_1 j_{\yo_1}} \dotsm \yqq_{i_\yL j_{\yo_\yL}}
\, \pr\bigl(\tland_\yj \yC^{(\yj)}_{j_\yj} \cond \yI\bigr),
\end{multline}
\begin{multline}
  \label{eq:new_ev_j_new}\tag{\ref{eq:new_ev_j}$'$}
  \pr(
\yR^{(\yo_{\yN + \yL})}_{i_{\yN + \yL}} \land \dotsb \land
\yR^{(\yo_{\yN + 1})}_{i_{\yN + 1}} 
\cond \yme \land
\yR^{(\yo_{\yN})}_{i_{\yN}} \land \dotsb \land \yR^{(\yo_{1})}_{i_{1}} 
\land \yI) 
={}
\\
\sum_{j_1, j_2, \dotsc} 
\Biggl[\tprod_{\yj=1}^{\yL}
\pr(\yR^{(\yo_\yj)}_{i_\yj} \cond \yM^{(\yo_\yj)} \land  
\yC^{(\yo_\yj)}_{j_\yj} \land \yI) \Biggr]
 \, \pr\bigl(\tland_\yj \yC^{(\yj)}_{j_{\yo_\yj}} \cond
\yR^{(\yo_{\yN})}_{i_{\yN}} \land \dotsb \land \yR^{(\yo_{1})}_{i_{1}}
\land
 \yI\bigr)
\equiv
{}\\
\sum_{j_1, j_2, \dotsc}
 \yqq_{i_1 j_{\yo_1}} \dotsm \yqq_{i_\yL j_{\yo_\yL}}
\, \pr\bigl(\tland_\yj \yC^{(\yj)}_{j_\yj} \cond
\yR^{(\yo_{\yN})}_{i_{\yN}} \land \dotsb \land \yR^{(\yo_{1})}_{i_{1}}
\land
 \yI\bigr),
\end{multline}
and
\begin{equation}
  \label{eq:upd_pr_Ck_new}\tag{\ref{eq:upd_pr_Cj}$'$}
\pr\bigl(\tland_\yj \yC^{(\yj)}_{j_\yj} \cond
\yR^{(\yo_{\yN})}_{i_{\yN}} \land \dotsb \land \yR^{(\yo_{1})}_{i_{1}}
\land
 \yI\bigr)
  =
  \frac{
 \yqq_{i_1 j_{\yo_1}} \dotsm \yqq_{i_\yN j_{\yo_\yN}}
\, \pr\bigl(\Land_\yj \yC^{(\yj)}_{j_\yj} \cond \yI\bigr)
  }{
\sum_{j_1, j_2, \dotsc} 
\yqq_{i_1 j_{\yo_1}} \dotsm \yqq_{i_\yN j_{\yo_\yN}}
\, \pr\bigl(\Land_\yj \yC^{(\yj)}_{j_\yj} \cond \yI\bigr)
  }.
\end{equation}
The plausibility distribution for the collection of
outcomes is therefore generally \emph{not} exchangeable
(the frequentist and the propensitor would say that the
events are not `identically distributed', though
independent).

The plausibilistic framework of the last generalisation is
used in non-equilibrium statistical
mechanics~\citep{jaynes1980c,jaynes1985b,dewar2003,dewar2005,dewar2005b}.
A generic circumstance $\yC^{(\yo)}_j$ represents a system's
being in a (`microscopic') state $j$ at the time $t_{\yo}$;
$R^{(\yo)}_i$ represents the obtainment of the $i$th outcome
of a (`macroscopic') measurement $\yM^{(\yo)}$ performed at
time $t_{\yo}$; the plausibilities $\pr(\yR^{(\yo)}_{i} \cond
\yM^{(\yo)} \land \yC^{(\yo)}_{j} \land \yI)$ are given by the
relevant physical theory for all $i,\yo,j$. We initially
assign, usually by means of maximum-entropy or
maximum-calibre principles, an initial plausibility
distribution $\pr({\textstyle\Land}_\yj \yC^{(\yj)}_{j_\yj}
\cond \yI)$ for all possible state dynamics $\Land_\yj
\yC^{(\yj)}_{j_\yj}$ that the system may follow. Then we
update, from the history of the observed measurement
outcomes, the plausibilities of the different dynamics and
thence those of unobserved outcomes.

Finally, a third generalisation, for which
see~\citep{portamana2007b,portamana2007}, is the introduction of multiple
kinds of measurements instead of a single one $\yM$. The resulting
representation, which should be easy for the reader to derive, would
correspond to de~Finetti's theorem for \emph{partial}
exchangeability~\citep{definetti1938,bernardoetal1994}.

\chapter{Conclusions}
\label{sec:concl}

Beside exchangeability, there is another point of view
from which the de~Finettians and the plausibilists can
approach the question of induction 
and thus also interpret and give meaning to the formulae
of the propensitors and some of their locutions like
`unknown probability' or `i.i.d.'. This point of view,
which we have named the `Laplace-Jaynes approach', is
based on an analysis of the particular situation of
interest into possible `circumstances'. These
`circumstances' are propositions that concern details of
an empirical nature; in particular, they are not and
cannot be statements about probabilities. They do not
necessarily involve notions of `cause' or `mechanism', but
can concern symmetries of the situation under study, or
consequences of the observed events. More generally, their
choice is fully `subjective', as are the plausibilities
assigned to them. Their main required property is that
when they are known they render irrelevant any
observational data for the purpose of assigning a
plausibility to unobserved events.

The Laplace-Jaynes approach can coexist with that based on
exchangeability, and can be a complement or an alternative
to the latter. It is particularly suited to problems in
natural philosophy, whose heart is the analysis of natural
phenomena into relevant circumstances (concerning
especially, but not exclusively, the notion of `cause').
It also allows --- without resorting to alternative
probability theories --- an interpretation, formalisation,
and quantification of feelings of `uncertainty' or
`instability' about some plausibility assignments.
Finally, this point of view is applicable in those
situations in which the maximum number of possible
observations is bounded (and, especially, small) and
infinite exchangeability cannot therefore be applied.

The Laplace-Jaynes approach is also straightforwardly
generalised to the case of more generic sets of
circumstances, the case of non-exchangeable plausibility
assignments (with applications in non-equilibrium
statistical mechanics), and the case in which different
kinds of measurements are present, usually approached by
partial exchangeability.

\begin{acknowledgements}
  \langitalian{PM non pu\`o ringraziare mai abbastanza
    Louise, Marianna, e Miriam per il loro continuo
    sostegno e amore.} Affectionate thanks also to the
  staff of the \langswedish{KTH Biblioteket}, especially
  the staff of the Forum Library --- Elisabeth Hammam,
  Ingrid Talman, Tommy Westergren, Elin Ekstedt, Thomas
  Hedbj\"orn, Daniel Larsson, Allan Lindqvist, Lina
  Lindstein, Yvonne Molin, Min Purroy Pei, Anders
  Robertsson --- for their patient and indefatigable work.
  Without them, research would hardly be possible. AM
  thanks Anders Karlsson for encouragement and advice.
\end{acknowledgements}



\newcommand{\bibpreamble}{Note: \texttt{arxiv} eprints are
  located at \url{http://arxiv.org/}.
}

\setlength{\bibsep}{0pt}

\newcommand{\bibfont}{\small}

\bibliography{bibliography}\end{document}

%% file: preamblemem.tex
\newcommand*\savesymbol[1]{%
  \expandafter\let\csname orig#1\expandafter\endcsname\csname#1\endcsname
  \expandafter\let\csname #1\endcsname\relax
}

\newcommand*\restoresymbol[2]{%
  \expandafter\global\expandafter\let\csname#1#2\expandafter\endcsname%
    \csname#2\endcsname
  \expandafter\global\expandafter\let\csname#2\expandafter\endcsname%
    \csname orig#2\endcsname
}

\usepackage[T1]{fontenc} 
\usepackage{textcomp}
\usepackage{amsmath}
\usepackage{amssymb}
\usepackage{amsxtra}
\usepackage[
]{babel}\FrenchItemizeSpacingfalse 
\newcommand{\langitalian}{\foreignlanguage{italian}}
\newcommand{\langswedish}{\foreignlanguage{swedish}}

\savesymbol{qedsymbol}

\usepackage{amsthm}

\restoresymbol{ams}{qedsymbol}
\newcommand{\QEE}
{\amsqedsymbol}
\newcommand{\qee}{\leavevmode\unskip\penalty9999 \hbox{}\nobreak\hfill
\quad\hbox{\QEE}}

\usepackage[dvipdfm]{graphicx}

\providecommand{\firstname}[1]{#1}
\providecommand{\surname}[1]{#1}

\providecommand{\usetxfonts}{}
\usetxfonts

\newcommand{\Alpha}{\mathrm{A}}
\let\varAlpha A
\newcommand{\Beta}{\mathrm{B}}
\let\varBeta B
\newcommand{\Epsilon}{\mathrm{E}}
\let\varEpsilon E
\newcommand{\Zeta}{\mathrm{Z}}
\let\varZeta Z
\newcommand{\Eta}{\mathrm{H}}
\let\varEta H
\newcommand{\Iota}{\mathrm{I}}
\let\varIota I
\newcommand{\Kappa}{\mathrm{K}}
\let\varKappa K
\newcommand{\Mu}{\mathrm{M}}
\let\varMu M
\newcommand{\Nu}{\mathrm{N}}
\let\varNu N
\newcommand{\Omicron}{\mathrm{O}}
\let\varOmicron O
\newcommand{\Rho}{\mathrm{P}}
\let\varRho P
\newcommand{\Tau}{\mathrm{T}}
\let\varTau T
\newcommand{\Chi}{\mathrm{X}}
\let\varChi X
\newcommand{\omicron}{\mathrm{o}}

\DeclareFontFamily{U}{egreek}{\skewchar\font'177}%
\DeclareFontShape{U}{egreek}{m}{n}{<-6>s*[1]eurm5<6-8>s*[1]eurm7 <8->s*[1]eurm10}{}%
\DeclareFontShape{U}{egreek}{m}{it}{<->s*[1]eurmo10}{}%
\DeclareFontShape{U}{egreek}{b}{n}{<-6>s*[1]eurb5 <6-8>s*[1]eurb7 <8->s*[1]eurb10}{}%
\DeclareFontShape{U}{egreek}{b}{it}{<->s*[1]eurbo10}{}%
\DeclareSymbolFont{egreeki}{U}{egreek}{m}{it}%
\SetSymbolFont{egreeki}{bold}{U}{egreek}{b}{it}
\DeclareMathSymbol{\epartial}{\mathalpha}{egreeki}{"40}

\DeclareMathSymbol{\ealpha}{\mathalpha}{egreeki}{"0B}
\DeclareMathSymbol{\ebeta}{\mathalpha}{egreeki}{"0C}
\DeclareMathSymbol{\egamma}{\mathalpha}{egreeki}{"0D}
\DeclareMathSymbol{\edelta}{\mathalpha}{egreeki}{"0E}
\DeclareMathSymbol{\eepsilon}{\mathalpha}{egreeki}{"0F}
\DeclareMathSymbol{\ezeta}{\mathalpha}{egreeki}{"10}
\DeclareMathSymbol{\eeta}{\mathalpha}{egreeki}{"11}
\DeclareMathSymbol{\etheta}{\mathalpha}{egreeki}{"12}
\DeclareMathSymbol{\eiota}{\mathalpha}{egreeki}{"13}
\DeclareMathSymbol{\ekappa}{\mathalpha}{egreeki}{"14}
\DeclareMathSymbol{\elambda}{\mathalpha}{egreeki}{"15}
\DeclareMathSymbol{\emu}{\mathalpha}{egreeki}{"16}
\DeclareMathSymbol{\enu}{\mathalpha}{egreeki}{"17}
\DeclareMathSymbol{\exi}{\mathalpha}{egreeki}{"18}
\DeclareMathSymbol{\eomicron}{\mathalpha}{egreeki}{"6F}
\DeclareMathSymbol{\epi}{\mathalpha}{egreeki}{"19}
\DeclareMathSymbol{\erho}{\mathalpha}{egreeki}{"1A}
\DeclareMathSymbol{\esigma}{\mathalpha}{egreeki}{"1B}
\DeclareMathSymbol{\etau}{\mathalpha}{egreeki}{"1C}
\DeclareMathSymbol{\eupsilon}{\mathalpha}{egreeki}{"1D}
\DeclareMathSymbol{\ephi}{\mathalpha}{egreeki}{"1E}
\DeclareMathSymbol{\echi}{\mathalpha}{egreeki}{"1F}
\DeclareMathSymbol{\epsi}{\mathalpha}{egreeki}{"20}
\DeclareMathSymbol{\eomega}{\mathalpha}{egreeki}{"21}
\DeclareMathSymbol{\evarepsilon}{\mathalpha}{egreeki}{"22}
\DeclareMathSymbol{\evartheta}{\mathalpha}{egreeki}{"23}
\DeclareMathSymbol{\evarpi}{\mathalpha}{egreeki}{"24}
\let\evarrho\erho 
\let\evarsigma\esigma
\let\evarkappa\ekappa
\DeclareMathSymbol{\evarphi}{\mathalpha}{egreeki}{"27}
\DeclareMathSymbol{\evarAlpha}{\mathalpha}{egreeki}{"41}
\DeclareMathSymbol{\evarBeta}{\mathalpha}{egreeki}{"42}
\DeclareMathSymbol{\evarGamma}{\mathalpha}{egreeki}{"00}
\DeclareMathSymbol{\evarDelta}{\mathalpha}{egreeki}{"01}
\DeclareMathSymbol{\evarEpsilon}{\mathalpha}{egreeki}{"45}
\DeclareMathSymbol{\evarZeta}{\mathalpha}{egreeki}{"5A}
\DeclareMathSymbol{\evarEta}{\mathalpha}{egreeki}{"48}
\DeclareMathSymbol{\evarTheta}{\mathalpha}{egreeki}{"02}
\DeclareMathSymbol{\evarIota}{\mathalpha}{egreeki}{"49}
\DeclareMathSymbol{\evarKappa}{\mathalpha}{egreeki}{"4B}
\DeclareMathSymbol{\evarLambda}{\mathalpha}{egreeki}{"03}
\DeclareMathSymbol{\evarMu}{\mathalpha}{egreeki}{"4D}
\DeclareMathSymbol{\evarNu}{\mathalpha}{egreeki}{"4E}
\DeclareMathSymbol{\evarXi}{\mathalpha}{egreeki}{"04}
\DeclareMathSymbol{\evarOmicron}{\mathalpha}{egreeki}{"4F}
\DeclareMathSymbol{\evarPi}{\mathalpha}{egreeki}{"05}
\DeclareMathSymbol{\evarRho}{\mathalpha}{egreeki}{"50}
\DeclareMathSymbol{\evarSigma}{\mathalpha}{egreeki}{"06}
\DeclareMathSymbol{\evarTau}{\mathalpha}{egreeki}{"54}
\DeclareMathSymbol{\evarUpsilon}{\mathalpha}{egreeki}{"07}
\DeclareMathSymbol{\evarPhi}{\mathalpha}{egreeki}{"08}
\DeclareMathSymbol{\evarChi}{\mathalpha}{egreeki}{"58}
\DeclareMathSymbol{\evarPsi}{\mathalpha}{egreeki}{"09}
\DeclareMathSymbol{\evarOmega}{\mathalpha}{egreeki}{"0A} 
\DeclareSymbolFont{egreekr}{U}{egreek}{m}{n}%
\SetSymbolFont{egreekr}{bold}{U}{egreek}{b}{n}
\DeclareMathSymbol{\epartialup}{\mathalpha}{egreekr}{"40}

\DeclareMathSymbol{\ealphaup}{\mathalpha}{egreekr}{"0B}
\DeclareMathSymbol{\ebetaup}{\mathalpha}{egreekr}{"0C}
\DeclareMathSymbol{\egammaup}{\mathalpha}{egreekr}{"0D}
\DeclareMathSymbol{\edeltaup}{\mathalpha}{egreekr}{"0E}
\DeclareMathSymbol{\eepsilonup}{\mathalpha}{egreekr}{"0F}
\DeclareMathSymbol{\ezetaup}{\mathalpha}{egreekr}{"10}
\DeclareMathSymbol{\eetaup}{\mathalpha}{egreekr}{"11}
\DeclareMathSymbol{\ethetaup}{\mathalpha}{egreekr}{"12}
\DeclareMathSymbol{\eiotaup}{\mathalpha}{egreekr}{"13}
\DeclareMathSymbol{\ekappaup}{\mathalpha}{egreekr}{"14}
\DeclareMathSymbol{\elambdaup}{\mathalpha}{egreekr}{"15}
\DeclareMathSymbol{\emuup}{\mathalpha}{egreekr}{"16}
\DeclareMathSymbol{\enuup}{\mathalpha}{egreekr}{"17}
\DeclareMathSymbol{\exiup}{\mathalpha}{egreekr}{"18}
\DeclareMathSymbol{\eomicronup}{\mathalpha}{egreekr}{"6F}
\DeclareMathSymbol{\epiup}{\mathalpha}{egreekr}{"19}
\DeclareMathSymbol{\erhoup}{\mathalpha}{egreekr}{"1A}
\DeclareMathSymbol{\esigmaup}{\mathalpha}{egreekr}{"1B}
\DeclareMathSymbol{\etauup}{\mathalpha}{egreekr}{"1C}
\DeclareMathSymbol{\eupsilonup}{\mathalpha}{egreekr}{"1D}
\DeclareMathSymbol{\ephiup}{\mathalpha}{egreekr}{"1E}
\DeclareMathSymbol{\echiup}{\mathalpha}{egreekr}{"1F}
\DeclareMathSymbol{\epsiup}{\mathalpha}{egreekr}{"20}
\DeclareMathSymbol{\eomegaup}{\mathalpha}{egreekr}{"21}
\DeclareMathSymbol{\evarepsilonup}{\mathalpha}{egreekr}{"22}
\DeclareMathSymbol{\evarthetaup}{\mathalpha}{egreekr}{"23}
\DeclareMathSymbol{\evarpiup}{\mathalpha}{egreekr}{"24}
\let\evarrhoup\erhoup 
\let\evarsigmaup\esigmaup
\let\evarkappaup\ekappaup
\DeclareMathSymbol{\evarphiup}{\mathalpha}{egreekr}{"27}
\DeclareMathSymbol{\eAlpha}{\mathalpha}{egreekr}{"41}
\DeclareMathSymbol{\eBeta}{\mathalpha}{egreekr}{"42}
\DeclareMathSymbol{\eGamma}{\mathalpha}{egreekr}{"00}
\DeclareMathSymbol{\eDelta}{\mathalpha}{egreekr}{"01}
\DeclareMathSymbol{\eEpsilon}{\mathalpha}{egreekr}{"45}
\DeclareMathSymbol{\eZeta}{\mathalpha}{egreekr}{"5A}
\DeclareMathSymbol{\eEta}{\mathalpha}{egreekr}{"48}
\DeclareMathSymbol{\eTheta}{\mathalpha}{egreekr}{"02}
\DeclareMathSymbol{\eIota}{\mathalpha}{egreekr}{"49}
\DeclareMathSymbol{\eKappa}{\mathalpha}{egreekr}{"4B}
\DeclareMathSymbol{\eLambda}{\mathalpha}{egreekr}{"03}
\DeclareMathSymbol{\eMu}{\mathalpha}{egreekr}{"4D}
\DeclareMathSymbol{\eNu}{\mathalpha}{egreekr}{"4E}
\DeclareMathSymbol{\eXi}{\mathalpha}{egreekr}{"04}
\DeclareMathSymbol{\eOmicron}{\mathalpha}{egreekr}{"4F}
\DeclareMathSymbol{\ePi}{\mathalpha}{egreekr}{"05}
\DeclareMathSymbol{\eRho}{\mathalpha}{egreekr}{"50}
\DeclareMathSymbol{\eSigma}{\mathalpha}{egreekr}{"06}
\DeclareMathSymbol{\eTau}{\mathalpha}{egreekr}{"54}
\DeclareMathSymbol{\eUpsilon}{\mathalpha}{egreekr}{"07}
\DeclareMathSymbol{\ePhi}{\mathalpha}{egreekr}{"08}
\DeclareMathSymbol{\eChi}{\mathalpha}{egreekr}{"58}
\DeclareMathSymbol{\ePsi}{\mathalpha}{egreekr}{"09}
\DeclareMathSymbol{\eOmega}{\mathalpha}{egreekr}{"0A}

\newcommand{\alleugreek}{%
\let\alpha\ealpha
\let\beta\ebeta
\let\gamma\egamma
\let\delta\edelta
\let\epsilon\eepsilon
\let\zeta\ezeta
\let\eta\eeta
\let\theta\etheta
\let\iota\eiota
\let\kappa\ekappa
\let\lambda\elambda
\let\mu\emu
\let\nu\enu
\let\xi\exi
\let\omicron\eomicron
\let\pi\epi
\let\rho\erho
\let\sigma\esigma
\let\tau\etau
\let\upsilon\eupsilon
\let\phi\ephi
\let\chi\echi
\let\psi\epsi
\let\omega\eomega
\let\varepsilon\evarepsilon
\let\vartheta\evartheta
\let\varpi\evarpi
\let\varrho\evarrho 
\let\varsigma\evarsigma
\let\varkappa\evarkappa
\let\varphi\evarphi
\let\varAlpha\evarAlpha
\let\varBeta\evarBeta
\let\varGamma\evarGamma
\let\varDelta\evarDelta
\let\varEpsilon\evarEpsilon
\let\varZeta\evarZeta
\let\varEta\evarEta
\let\varTheta\evarTheta
\let\varIota\evarIota
\let\varKappa\evarKappa
\let\varLambda\evarLambda
\let\varMu\evarMu
\let\varNu\evarNu
\let\varXi\evarXi
\let\varOmicron\evarOmicron
\let\varPi\evarPi
\let\varRho\evarRho
\let\varSigma\evarSigma
\let\varTau\evarTau
\let\varUpsilon\evarUpsilon
\let\varPhi\evarPhi
\let\varChi\evarChi
\let\varPsi\evarPsi
\let\varOmega\evarOmega 
\let\alphaup\ealphaup
\let\betaup\ebetaup
\let\gammaup\egammaup
\let\deltaup\edeltaup
\let\epsilonup\eepsilonup
\let\zetaup\ezetaup
\let\etaup\eetaup
\let\thetaup\ethetaup
\let\iotaup\eiotaup
\let\kappaup\ekappaup
\let\lambdaup\elambdaup
\let\muup\emuup
\let\nuup\enuup
\let\xiup\exiup
\let\omicronup\eomicronup
\let\piup\epiup
\let\rhoup\erhoup
\let\sigmaup\esigmaup
\let\tauup\etauup
\let\upsilonup\eupsilonup
\let\phiup\ephiup
\let\chiup\echiup
\let\psiup\epsiup
\let\omegaup\eomegaup
\let\varepsilonup\evarepsilonup
\let\varthetaup\evarthetaup
\let\varpiup\evarpiup
\let\varrhoup\evarrhoup 
\let\varsigmaup\evarsigmaup
\let\varkappaup\evarkappaup
\let\varphiup\evarphiup
\let\Alpha\eAlpha
\let\Beta\eBeta
\let\Gamma\eGamma
\let\Delta\eDelta
\let\Epsilon\eEpsilon
\let\Zeta\eZeta
\let\Eta\eEta
\let\Theta\eTheta
\let\Iota\eIota
\let\Kappa\eKappa
\let\Lambda\eLambda
\let\Mu\eMu
\let\Nu\eNu
\let\Xi\eXi
\let\Omicron\eOmicron
\let\Pi\ePi
\let\Rho\eRho
\let\Sigma\eSigma
\let\Tau\eTau
\let\Upsilon\eUpsilon
\let\Phi\ePhi
\let\Chi\eChi
\let\Psi\ePsi
\let\Omega\eOmega
}

\newcommand{\unambeugreek}{%
\let\omicron\eomicron
\let\varAlpha\evarAlpha
\let\varBeta\evarBeta
\let\varEpsilon\evarEpsilon
\let\varZeta\evarZeta
\let\varEta\evarEta
\let\varIota\evarIota
\let\varKappa\evarKappa
\let\varMu\evarMu
\let\varNu\evarNu
\let\varOmicron\evarOmicron
\let\varRho\evarRho
\let\varTau\evarTau
\let\varChi\evarChi
\let\omicronup\eomicronup
\let\Alpha\eAlpha
\let\Beta\eBeta
\let\Epsilon\eEpsilon
\let\Zeta\eZeta
\let\Eta\eEta
\let\Iota\eIota
\let\Kappa\eKappa
\let\Mu\eMu
\let\Nu\eNu
\let\Omicron\eOmicron
\let\Rho\eRho
\let\Tau\eTau
\let\Chi\eChi
}

\usepackage{bm}
\providecommand{\piup}{\epiup}
\providecommand{\deltaup}{\edeltaup}
\providecommand{\zetaup}{\ezetaup}
\providecommand{\varepsilonup}{\evarepsilonup}

\providecommand{\coloneqq}{\mathrel{\mathop:}=}
\providecommand{\eqqcolon}{=\mathrel{\mathop:}}

\newcommand{\delt}{\deltaup}
\newcommand{\di}{\mathrm{d}}

\DeclareSymbolFont{AMSb}{U}{msb}{m}{n}
\DeclareMathSymbol{\CC}{\mathalpha}{AMSb}{"43}
\DeclareMathSymbol{\RR}{\mathalpha}{AMSb}{"52}
\DeclareMathSymbol{\NN}{\mathalpha}{AMSb}{"4E}
\DeclareMathSymbol{\ZZ}{\mathalpha}{AMSb}{"5A}
\DeclareMathSymbol{\QQ}{\mathalpha}{AMSb}{"51}

\newcommand{\defd}{\coloneqq}
\newcommand{\defs}{\eqqcolon}
\newcommand{\Land}{\bigwedge}

\newcommand{\cond}
{\mathpunct{|}}
\newcommand{\bigcond}{\mathpunct{\big|}}
\newcommand{\st}{\mid}
\newcommand{\with}{\colon}

\renewcommand{\ge}{\geqslant}
\DeclareMathDelimiter{\lclose}{\mathopen}{operators}{"5B}{largesymbols}{"02}
\DeclareMathDelimiter{\rclose}{\mathclose}{operators}{"5D}{largesymbols}{"03}
\DeclareMathDelimiter{\lopen}{\mathopen}{operators}{"5D}{largesymbols}{"03}
\DeclareMathDelimiter{\ropen}{\mathclose}{operators}{"5B}{largesymbols}{"02}
\newcommand{\clcl}[1]{\lclose#1\rclose}

\newcommand{\opop}[1]{\lopen#1\ropen}

\newcommand{\set}[1]{\{#1\}}
\DeclareMathOperator{\pr}{P}
\DeclareMathOperator{\pf}{p}
\newcommand{\prop}[1]{{\textnormal{`#1'}}}
\newcommand{\sect}{\S}
\newcommand{\sects}{\S\S}
\newcommand{\chap}{ch.}%
\newcommand{\eqn}{eq.}%
\newcommand{\eqns}{eqs.}%

\theoremstyle{remark}
\newtheorem{remark}{Remark}
\theoremstyle{definition}

\newcommand{\etc}{{etc.}}
\newcommand{\ie}{{i.e.}}
\newcommand{\Ie}{{I.e.}}

\newcommand{\eg}{{e.g.}}
\newcommand{\viz}{{viz.}}
\newcommand{\cf}{{cf.}}
\newcommand{\Cf}{{Cf.}}

\newcommand{\etal}{{et al.}}

\newcommand{\bd}{\hspace{0pt}}%

\newcommand{\labelbis}[1]{\tag*{(\ref{#1})$_\text{r}$}}

\providecommand{\href}[2]{#2}